\documentclass[aps]{revtex4}
\usepackage{epsfig, amssymb, latexsym, amsfonts, amsmath, amsthm}

\newcommand{\PP}{ {\mathbb{P}} }
\newcommand{\KI}{ {\mathcal{K}} }
\newcommand{\QU}{ {\mathcal{Q}} }
\newcommand{\PU}{ {\mathcal{P}} }
\newcommand{\G}{ {\mathsf{G}} }
\newcommand{\T}{ {\mathsf{T}} }
\newcommand{\F}{ {\mathsf{F}} }
\newcommand{\twz}{ {\text{tw2}} }

\def \d{{\mathrm{d}}}
\def \D{{\mathrm{D}}}
\def \pd{\partial}
\newcommand{\M}{\mu}

\newcommand {\abs}[1]{\left| #1 \right|}

\newcommand {\kln}[1]{\left( #1 \right)}
\newcommand {\Kln}[1]{\bigl( #1 \bigr)}
\newcommand {\KLn}[1]{\Bigl( #1 \Bigr)}

\newcommand {\klno}[1]{\left( #1 \right.}
\newcommand {\Klno}[1]{\bigl( #1 \bigr.}

\newcommand {\okln}[1]{\left. #1 \right)}
\newcommand {\oKln}[1]{\bigl. #1 \bigr)}

\newcommand {\kls}[1]{\left\{ #1 \right\}}

\newcommand {\KLS}[1]{\biggl\{ #1 \biggr\}}

\newcommand {\klso}[1]{\left\{ #1 \right.}

\newcommand {\KLSo}[1]{\biggl\{ #1 \biggr.}
\newcommand {\KLSSo}[1]{\Biggl\{ #1 \Biggr.}

\newcommand {\oKLS}[1]{\biggl. #1 \biggr\}}
\newcommand {\oKLSS}[1]{\Biggl. #1 \Biggr\}}

\newcommand {\kle}[1]{\left[ #1 \right]}

\newcommand {\KLe}[1]{\Bigl[ #1 \Bigr]}
\newcommand {\KLE}[1]{\biggl[ #1 \biggr]}

\newcommand {\KLEo}[1]{\biggl[ #1 \biggr.}
\newcommand {\KLEEo}[1]{\Biggl[ #1 \Biggr.}

\newcommand {\oKLE}[1]{\biggl. #1 \biggr]}
\newcommand {\oKLEE}[1]{\Biggl. #1 \Biggr]}

\newcommand {\matel}[3]{\left< #1 \left|\; #2\, \right| #3 \right>}
\newcommand {\Matel}[3]{\bigl< #1 \bigl|\; #2\, \bigr| #3 \bigr>}
\newcommand {\MAtel}[3]{\Bigl< #1 \Bigl|\; #2\, \Bigr| #3 \Bigr>}

\newcommand {\ket}[1]{\left|\; #1\, \right>}

\newcommand {\bra}[1]{\left<\; #1\, \right|}

\newcommand {\kom}[2]{\kle{#1\;,\,#2}}

\newcommand {\PIPe}[1]{\biggl|_{#1} \biggr.}

\newcommand {\R} {{\mathbb{R}}}

\newcommand {\Id} {{\mathbb{I}}}
\newcommand {\im}{{\text{i}}}
\newcommand {\e}{{\text{e}}}

\newcommand {\Fe}[1]{#1 \!\!\! /\,}

\newcommand {\lra} {\leftrightarrow}

\newcommand {\Ra} {\longrightarrow}

\newcommand {\ra} {\rightarrow}

\newcommand {\ix} {\int {\mathrm{d}}^4 \! x}

\newcommand {\tx} {\tilde x}

\begin{document}

\sloppy
\begin{flushleft} DESY 01-107 \hfill {\tt hep-ph/0108095}
\\ NTZ 11/2001 \\ August 2001 \end{flushleft}

\title{On the structure of the virtual Compton amplitude
with additional final-state meson in the extended Bjorken region}

\author{ Johannes Bl\"umlein }
\email{ johannes.bluemlein@desy.de }
\affiliation{ Deutsches Elektronen-Synchrotron, DESY-Zeuthen,
Platanenallee 6, D-15735 Zeuthen, Germany }

\author{ J\"org Eilers }
\email{ eilers@itp.uni-leipzig.de }
\affiliation{ Center for Theoretical Studies and Institute of
Theoretical Physics, Leipzig University, Augustusplatz~10,
D-04109~Leipzig, Germany}

\author{ Bodo Geyer }
\email{ geyer@itp.uni-leipzig.de }
\affiliation{ Center for Theoretical Studies and Institute of
Theoretical Physics, Leipzig University, Augustusplatz~10,
D-04109~Leipzig, Germany}

\author{ Dieter Robaschik }
\email{ drobasch@physik.tu-cottbus.de }
\affiliation{BTU Cottbus, Fakult\"at 1, Postfach 101344,
D-03013~Cottbus, Germany }

\date{\today}

%!p###################################################################
\begin{abstract}

\vspace*{0.5cm}

\noindent
Using the framework of the non-local light-cone expansion a
systematic study is performed for the structure of the twist-2
contributions to the virtual Compton amplitude in polarized
deep-inelastic non-forward scattering for general nucleon spin with an
additional scalar meson in the final state. A useful kinematic
parameterization allowing for appropriate triple-valued off-forward
parton distribution amplitudes is given. One-variable amplitudes being
adapted to the fixed parameters of the extended Bjorken region are
introduced by decomposing the Compton amplitude into collinear and
non-collinear components. These amplitudes obey Wandzura-Wilczek and
Callan-Gross like relations. The evolution equations for all the
distribution amplitudes are determined showing that the additional
meson momentum does not appear in the evolution kernels. The
generalization to $n$ outgoing mesons is given.

\vspace*{0.5cm}

\noindent
PACS: 24.85.+p,  13.88.+e, 11.30.Cp\\
Keywords: Twist decomposition,
Nonlocal light-cone operators,
Tensor harmonic polynomials,
Multivalued distribution amplitude,
Extended Bjorken region

\end{abstract}
%!p###################################################################

\maketitle

%%%%%%%%%%%%%%%%%%%%%%%%%%%%%%%%%%%%%%%%%%%%%%%%%%%%%%%%%%%%%%%%%%%%%%
\section{Introduction}
\setcounter{equation}{0}
%%%%%%%%%%%%%%%%%%%%%%%%%%%%%%%%%%%%%%%%%%%%%%%%%%%%%%%%%%%%%%%%%%%%%%

\noindent
Compton scattering of a virtual photon off a hadron,
%!p___________________________________________________________________
\begin{equation}
\label{ohne_meson}
 {\gamma^*}(q_1) + {\mathrm H}(p_1) \Ra {\gamma^*}(q_2) +{\mathrm
H}(p_2),
\end{equation}
%!p___________________________________________________________________
is an important process in Quantum Chromodynamics. This general
process covers a series of different reactions through which a variety
of inclusive informations on the short--distance structure of nucleons
become accessible at large space--like virtualities. It is also
closely connected to the spin problem of the nucleon. The case of
forward scattering $p_1 = p_2 = p$ describes deep inelastic scattering
(DIS) off unpolarized or polarized targets which is widely discussed
in the literature, see e.g.~\cite{DIS,MUTA}, and $p_1 \neq p_2$
corresponds to the generic (ordinary) non-forward virtual Compton
scattering~\cite{MRGHD,CS,BGR97,BGR99,BR00}, for recent reviews
see~\cite{REV1}. In the special case
$q_2^2 = 0$, i.e.,~when the outgoing photon is real,
this process is called deeply virtual Compton scattering (DVCS).

Experimental results on polarized and unpolarized (deeply virtual) Compton
scattering were reported in~\cite{EXP0,EXP1,EXP2,EXP3,EXP4}. The kinematic domain of some
of these investigations is bound to rather low values of $Q^2$.
Experimentally the final state in deep--inelastic non--forward
scattering contains aside the (virtual) photon and final--state
hadron, Eq.~(1.1), a series of other hadrons, which even may emerge at
the amplitude--level. The latter process is much more likely
and of greater practical
importance than that
of a single isolated hadron
$\gamma_1^* + H_1 \rightarrow \gamma^*_2 +H_2$, which was studied
before~\cite{MRGHD,CS,BGR97,BGR99,BR00}.

In this paper we extend the description given in the ordinary
non-forward case in Refs.~\cite{BGR97,BGR99,BR00} to physical processes with an
outgoing scalar meson,
%!p___________________________________________________________________
\begin{equation}
\label{mit_meson}
 {\gamma^*}(q_1) + {\mathrm H}(p_1) \Ra {\gamma^*}(q_2) +{\mathrm
H}(p_2) + {\mathrm M}(k) \; ,
\end{equation}
%!p___________________________________________________________________
to investigate which of the properties derived in
Refs.~\cite{BGR97,BGR99,BR00} remain valid in the more general
situation and which are changing to account for more realistic
experimental situations which allow for additional studies of
distribution functions emerging in non--forward scattering
\footnote{Similarly to the case of deep--inelastic non--forward
scattering multiple hadron final states also emerge in deep--inelastic
diffractive scattering~\cite{EXPDIFFR} along the diffractive final
state proton, which is assumed an isolated particle in the ideal case,
c.f. Ref.~\cite{BR01}.}.

If one looks for a diagrammatic representation of the amplitude for the
process (\ref{mit_meson}) then even in the special kinematics
of the Bjorken region for $\kln{p_2 + k - p_1}^2$ being small, there appear
different production mechanisms.
However, if one has in mind $p_2^2 = M^2, \, k^2 =m^2$
and $p_2.k \ll M^2$ then soft processes between the two final
particles and the incoming particle $\ket{p_1}$ are essential and
we have to use a generalized distribution amplitude
$\matel{p_2,k}{O}{p_1}$. For other cases one may try to make models
which use wave functions of the nucleon $\bra{p_2}$ and the meson
$\bra{k}$.

The Compton amplitude for the process (\ref{mit_meson}) is given by
%!p___________________________________________________________________
\begin{equation}
\label{Compton_amp}
 T_{\mu\nu}\kln{P_+,P_-,k;q} = \im \ix \; \e^{\im qx} \,
\matel{p_2,S_2;k,0}{R \, T\kle{J_\mu\kln{\frac{x}{2}}
J_\nu\kln{-\frac{x}{2}}{\cal S}} }{p_1,S_1}
\end{equation}
%!p___________________________________________________________________
where
%!p___________________________________________________________________
\begin{eqnarray}
 P_\pm \equiv P_f \pm P_i = p_2 \pm p_1 + k, \quad q =
\hbox{$\frac{1}{2}$} \kln{q_1 + q_2} \, ,
\end{eqnarray}
%!p___________________________________________________________________
and $k$ are chosen as independent kinematic variables. As usual
$p_1\kln{p_2}$ and $q_1\kln{q_2}$ denote the four-momenta of the
incoming (outgoing) nucleons and photons, respectively. $S_1$ and
$S_2$ are the spins of these nucleons and $k$ is the momentum of the
outgoing scalar meson. $p_\pm = p_2 \pm p_1$ denote the independent
kinematic variables in the ordinary non-forward case. In principle,
this set of kinematic variables can be extended to the case of $n$
outgoing (scalar) mesons of momenta $k_i,\,i=1,\ldots,n$ with
$P_f = p_2 + \sum_i k_i$.

In ordinary non-forward Compton scattering the \textit{generalized
Bjorken region} is defined by the conditions
%!p___________________________________________________________________
\begin{equation}
\label{int_bjorken1}
 \nu \equiv q p_+ \Ra \infty \qquad {\text{and}} \qquad Q^2 \equiv
-q^2 \Ra \infty \, ,
\end{equation}
%!p___________________________________________________________________
keeping the variables
%!p___________________________________________________________________
\begin{equation}
\label{int_bjorken2}
 \xi = - \frac{q^2}{q p_+} \qquad \text{and} \qquad \eta = \frac{q
p_-}{q p_+} = \frac{q_1^2 - q_2^2}{2\nu}
\end{equation}
%!p___________________________________________________________________
fixed. This definition has to be extended now by taking into account
the additional momentum $k$. We define the \textit{extended Bjorken
region} for light-cone dominated QCD processes by the conditions
(\ref{int_bjorken1}) keeping the following three variables
%!p___________________________________________________________________
\begin{equation}
\label{scalen_variable}
 \xi = - \frac{q^2}{q P_+} \, , \qquad \eta = \frac{q P_-}{q P_+}
\qquad \text{and} \qquad \chi = \frac{q k }{q P_+}
\end{equation}
%!p___________________________________________________________________
fixed. This can be extended furthermore to the case of $n$ mesons by
the obvious generalization of $P_+$ and keeping
$\chi_i = (qk_i)/(qP_+)$ fixed. In the limit $k_i \ra 0$, these
definitions reduce to the usual \textit{generalized Bjorken region}.

For completeness, we list the different kinematic domains for
forward and general non-forward processes and the related scaling
variables. These domains are distinguished as follows:
%!p<><><><><><><><><><><><><><><><><><><><><><><><><><><><><><><><><><
\begin{itemize}
%!p<><><><><><><><><><><><><><><><><><><><><><><><><><><><><><><><><><
\item
{\textit{Bjorken region} for forward scattering; fixed quantity:
$\xi$ (or $x_B=-q_1^2/2q_1.p_1$)}
%!p<><><><><><><><><><><><><><><><><><><><><><><><><><><><><><><><><><
\item
{\textit{Generalized Bjorken region} for ordinary non-forward
scattering, including DVCS kinematics for $q_2^2 = 0$; fixed
quantities: $\xi$ and $\eta$. In the special case of DVCS
$\xi = - \eta$ holds.}
%!p<><><><><><><><><><><><><><><><><><><><><><><><><><><><><><><><><><
\item
{\textit{Extended Bjorken region} for non-forward scattering with a
single outgoing scalar meson; fixed quantities: $\xi, \eta$ and
$\chi$. }
%!p<><><><><><><><><><><><><><><><><><><><><><><><><><><><><><><><><><
\item
{$n$-\textit{Extended Bjorken region} for non-forward scattering
with $n$ outgoing scalar mesons; fixed quantities: $\xi, \eta$ and
$\chi_1,\dots,\chi_n$.}
%!p<><><><><><><><><><><><><><><><><><><><><><><><><><><><><><><><><><
\end{itemize}
%!p<><><><><><><><><><><><><><><><><><><><><><><><><><><><><><><><><><
By conservation of momentum,
%!p___________________________________________________________________
\begin{equation}
 p_1 + q_1 = p_2 + q_2 + \hbox{$\sum_i$} \, k_i \, ,
\end{equation}
%!p___________________________________________________________________
one can show that $\xi$ and $\eta$ possess the same interpretation
as in the ordinary non-forward case. Especially, also the relation
$\eta = \kln{q_1^2 - q_2^2}/2\nu$ holds for processes including the
outgoing mesons. The variables $\chi_i$ describe the momentum
fractions of the mesons $k_i$ in the infinite momentum frame defined
by the momentum $P_+$. In the following we restrict the consideration
to the case of one additional final-state meson and return to the case
of $n$-mesons in Section~\ref{gen}.

It is important to remark that all physical processes mentioned
above are distinguished only by taking different matrix elements of
the same renormalized operator~\footnote{As has been shown
recently in Ref.~\cite{BR01}, also the case of deep inelastic diffractive
scattering can be described following similar lines.}, namely the
renormalized ($R$) time-ordered ($T$) product
%!p___________________________________________________________________
\begin{equation}
\label{int_input}
 \widehat T_{\mu\nu}(x) \equiv R \, T\kle{J_\mu\kln{\frac{x}{2}} \,
J_\nu\kln{-\frac{x}{2}} {\cal S} },
\end{equation}
%!p___________________________________________________________________
where $J_\mu(x)= :\!\bar\psi(x)\gamma_\mu\psi(x)\!:$ denotes the
hadronic current and $\cal S$ is the renormalized $S-$matrix. Near the
light-cone, $x^2 \ra 0,$ this operator will be decomposed via the
non-local operator product expansion \cite{AZ78} into a series of
non-local light-ray operators and the corresponding coefficient
functions~:
%!p___________________________________________________________________
\begin{eqnarray}
\label{int_lce}
\nonumber
 \widehat T_{\mu\nu}(x)
&=&
 \!\! \int_{-1}^1 \d^2 \! \kappa \;\,
C_{\Gamma\mu\nu}\!\kln{x^2,\kappa_i\tx;\mu^2} \, \! R \, T \!
\kle{O^\Gamma\!\kln{\kappa_1\tx,\kappa_2\tx}{\cal S}}
\\
&&
 + \quad \text{higher order terms} \, .
\end{eqnarray}
%!p___________________________________________________________________
The coefficient functions are singular on the light-cone. They are
entire analytic functions with respect to $\kappa_i \tx$ resulting in
a restricted integration range $-1\leq \kappa_i \leq 1$. The
unrenormalized light-ray operators in this expansion are given by
%!p___________________________________________________________________
\begin{equation}
 O^\Gamma \! \kln{\kappa_1\tx,\kappa_2\tx} = \;\;
:\bar\psi\kln{\kappa_1\tx} \; \Gamma \; U\kln{\kappa_1\tx,\kappa_2\tx}
\, \psi\kln{\kappa_2\tx} :
\end{equation}
%!p___________________________________________________________________
with some specified $\Gamma$-structure of Dirac matrices and the
usual path-ordered phase factor,
%!p___________________________________________________________________
\begin{equation}
 U\kln{\kappa_1\tx,\kappa_2\tx} = \text{P} \; \exp\kln{-\im \, g
\int_{\kappa_2}^{\kappa_1} \d\tau \; \tx^\mu A_\mu\kln{\tau\tx} } \, ,
\end{equation}
%!p___________________________________________________________________
ensuring gauge invariance. Here $A_\mu$ is the gluon field, $g$ the
strong coupling constant and $\tx$ is a light-like vector depending on
$x$ via a non-null subsidiary four-vector $\zeta$,
%!p___________________________________________________________________
\begin{equation}
 \tx = x - \frac{\zeta}{\zeta^2}
\kln{\kln{x\zeta}-\sqrt{\kln{x\zeta}^2 - x^2\zeta^2}} \quad \;
\text{with} \quad \; \zeta^2 > 0.
\end{equation}
%!p___________________________________________________________________

In Eq.~(\ref{int_lce}) the flavor structure has been suppressed.
Eventually, in the singlet case, also operators containing the gluon
field strength $F_{\mu\nu}$ and their dual $\widetilde F_{\mu\nu}$
have to be taken into account. The contributions which contain four or
more (anti)quark fields will be denoted by `higher order terms',
possibly also together with (powers of) the gluon field strength, etc.
By construction, the non-local light-cone expansion, depending on the
order of terms being taken into account, leads to a
(sub)asymptotically relevant part and a well-defined remainder being
less singular, see Ref.~\cite{MUTA}.

Taking matrix elements of the operators $O^\Gamma$ and performing a
Fourier transformation in the expression (\ref{Compton_amp}) leads to
the physically interesting non-perturbative distribution amplitudes.
Because the operational input (\ref{int_input}) for these amplitudes
is the same for the different processes described above, the evolution
equations of these amplitudes are determined by the renormalization
group equations of the operators $O^\Gamma$ and, therefore, also by
the same anomalous dimensions.

This paper is organized as follows. In Section~\ref{str} we discuss
the quark-antiquark operator (\ref{str_oprab}) as operational input
for the (extended) Compton amplitude. This allows one to use the known
anomalous dimensions of this operator to
write down the evolution equations for the relevant distribution
amplitudes used in the process including an outgoing meson. The
evolution kernel required for these evolution equations is determined
by the anomalous dimensions of the quark-antiquark operator and
further computations of Feynman diagrams are not necessary, at least
at one--loop order. Furthermore this operator possesses a known twist
decomposition given in Refs.~\cite{GLR99,GLR01}. In this paper we only
determine the twist-2 part of the Compton amplitude in the simply
\textit{extended Bjorken region} which generalizes the twist-2
representation known for forward and ordinary non-forward scattering.

In Section~\ref{twi} the twist decomposition of the matrix elements
and the necessary decomposition into suitable kinematic factors is
performed thereby using the hadron equations of motion. This section
also includes the definition of the distribution amplitudes used in
the process considered.

In Section~\ref{com} the twist-2 part of the Compton amplitude is
calculated. It will be shown that, in the extended kinematic region,
the amplitude depends on the three scaling variables $\xi,\eta$ and
$\chi$ and is given by \textit{triple-valued distributions}.

In Section~\ref{int_relations} we extract integral relations
contained in the Compton amplitude and Section~\ref{evo} includes the
evolution equations obeyed by the distribution amplitudes.

In Section~VII we add some remarks on generic properties of the
results obtained in the preceding sections and consider the case of
$n$ outgoing mesons; Section~VIII contains the conclusions.

Appendix~A contains the projections of the twist--2 operators on the
light cone. In the Appendix~\ref{helicity projection} we construct the
helicity basis for both photons and calculate all the helicity
amplitudes. These projections show in explicit form that the process
(\ref{mit_meson}) is current conserving on the level of the
$S$--matrix.

%%%%%%%%%%%%%%%%%%%%%%%%%%%%%%%%%%%%%%%%%%%%%%%%%%%%%%%%%%%%%%%%%%%%%%
\section{Operator Structure}
\label{str}
\setcounter{equation}{0}
%%%%%%%%%%%%%%%%%%%%%%%%%%%%%%%%%%%%%%%%%%%%%%%%%%%%%%%%%%%%%%%%%%%%%%

\noindent
In this section we discuss the operator structure of the Compton
amplitude which will be used in the following sections. For brevity we
discuss only the flavor non-singlet case and drop all flavor
structures in the operators. The construction for the flavor singlet
case is to be carried out similarly.

As was mentioned above, the operator input for the Compton amplitude
$T_{\mu\nu}(P_\pm,k;q)$ is given by the renormalized time-ordered
product of two electromagnetic currents, Eq.~(\ref{int_input}).
Obviously, in lowest order in the coupling constant the $S$-matrix can
be set equal to one. Then, applying the Wick-Theorem to this
time-ordered product and approximating the quark propagator near the
light-cone by
%!p___________________________________________________________________
\begin{eqnarray}
\label{propagator}
 S(x,m)
&\equiv&
 \kln{ \im \, \Fe{\partial} + m } \Delta\kln{x,m}
\\
\nonumber
&\approx&
 - \frac{\Fe{x}}{2\pi^2 \kln{x^2 - \im \epsilon}^2} -
\frac{m}{8\pi^2\kln{x^2 - \im \epsilon}}\kln{2\im + {m}\, \Fe{x}} +
\frac{m^3}{128 \pi^2} \ln\kln{-x^2 + \im \epsilon} \kln{8\im + {m}\,
\Fe{x}}
\end{eqnarray}
%!p___________________________________________________________________
one obtains the following description of the operator
$\widehat T_{\mu\nu}(x)$:
%!p___________________________________________________________________
\begin{eqnarray}
\label{str_wick}
 T\kle{J_\mu\kln{\frac{x}{2} }J_\nu\kln{-\frac{x}{2} }}
&\approx&
 -x^\alpha\,\kle{ \frac{1}{2\pi^2\kln{x^2 - \im \epsilon}^2} +
\frac{m^2}{8\pi^2\kln{x^2 - \im \epsilon}} - \frac{m^4}{128\pi^2}
\ln\kln{-x^2 + \im \epsilon} }
\\
\nonumber
&&
 \qquad \qquad \times \KLn{ :\!\bar\psi\kln{x/2}\,\gamma_\mu
\gamma_\alpha \gamma_\nu \,\psi\kln{-x/2}\!: -:\!\bar\psi\kln{-x/2}\,
\gamma_\nu \gamma_\alpha \gamma_\mu\,\psi\kln{x/2}\!:}
\\
\nonumber
&&
 - \; \im\,m\,\kle{ \frac{1}{4\pi^2\kln{x^2 - \im \epsilon} } -
\frac{m^2 }{16\pi^2} \ln\kln{-x^2 + \im \epsilon} }
\\
\nonumber
&&
 \qquad \qquad \times \KLEo{\KLn{:\!\bar\psi\kln{x/2} \,
\sigma_{\mu\nu} \, \psi\kln{-x/2}\! : - :\!\bar\psi\kln{-x/2} \,
\sigma_{\mu\nu} \, \psi\kln{x/2}\! :}}
\\
\nonumber
&&
 \qquad \qquad \qquad + \oKLE{\;g_{\mu\nu} \,\KLn{:\!\bar\psi\kln{x/2}
\; \psi\kln{-x/2}\! : + \, :\!\bar\psi\kln{-x/2} \, \psi\kln{x/2}\!:
}}
\\
\nonumber
&&
 + \, :\bar\psi\kln{x/2} \gamma_\mu \psi\kln{x/2} \bar\psi\kln{-x/2}
\gamma_\nu \psi\kln{-x/2}: \; + \; \frac{2x_\mu x_\nu -
g_{\mu\nu}x^2}{\pi^4\, \kln{x^2 - \im \epsilon}^4}~.
\end{eqnarray}
%!p___________________________________________________________________
The renormalization symbol $R$ is suppressed in the notation. The
four-quark operator
$\bar\psi \gamma_\mu \psi \, \bar\psi \gamma_\nu \psi$ has no singular
coefficient on the light-cone and is of minimal twist-4 and the last
term stems from disconnected diagrams. Both terms are therefore
discarded, see Ref.~\cite{MUTA}.

The quark mass terms resulting from the mass
dependent part of the operator $\kln{\im\Fe{\partial} + m}$ are less
singular and will also be omitted. In fact, the lowest twist contribution
of all quark mass terms is contained in the operator
%!p___________________________________________________________________
\begin{equation}
 m \KLn{:\!\bar\psi\kln{x/2}\,\im\,\sigma_{\mu\nu}\,\psi\kln{-x/2}\!:
- :\!\bar\psi\kln{-x/2}\,\im\,\sigma_{\mu\nu}\,\psi\kln{x/2}\!:}
\end{equation}
%!p___________________________________________________________________
with $\sigma_{\mu\nu} = (\im/2)[\gamma_\mu,\,\gamma_\nu]$ and is of
twist-3.

The first term of the expansion (\ref{str_wick}) is of main
importance, because it contains the leading light-cone singularity and
its minimal twist contribution is of twist-2. It also contains terms
of higher twist (trace terms) and quark mass terms resulting
from the mass dependence of the scalar propagator $\Delta\kln{x.m}$.
These terms are also less singular on the light-cone.

Using the standard relations
%!p___________________________________________________________________
\begin{eqnarray}
\label{str_zer}
 \gamma_\mu\gamma_\alpha \gamma_\nu
&=&
 S_{\mu\alpha\nu\beta}\, \gamma^\beta + \im \,
\varepsilon_{\mu\alpha\nu\beta}\, \gamma^5 \gamma^\beta\ ,
\\
\label{SYM}
 S_{\mu\nu|\alpha\beta}
&\equiv&
 S_{\mu\alpha\nu\beta}
\\
\nonumber
& \equiv&
 \KLn{g_{\mu\alpha} \, g_{\nu\beta} + g_{\nu\alpha} \, g_{\mu\beta} -
g_{\mu\nu} \, g_{\alpha\beta} },
\end{eqnarray}
%!p___________________________________________________________________
where we indicate that $S_{\mu\alpha\nu\beta}$ is symmetric in
$\mu\nu$ and $\alpha\beta$, and noticing that the operator
(\ref{str_wick}) is dominated by the light-cone singularity,
$x^2 \approx 0$, one arrives at
%!p___________________________________________________________________
\begin{eqnarray}
\label{approximation_used}
 \widehat T_{\mu\nu}(x) &\approx & \frac{\im \;
\tx^\alpha}{\pi^2\kln{x^2 - \im \epsilon}^2}\; \!\!
\klno{S_{\mu\alpha\nu\beta} \; R\,T \!
\kle{O^\beta\kln{\frac{\tx}{2},-\frac{\tx}{2}}{\cal S}}} \okln{ + \;
\im \, \varepsilon_{\mu\alpha\nu\beta} \; R\,T \! \kle{O^{5 \,\beta}
\kln{\frac{\tx}{2},-\frac{\tx}{2}}{\cal S}}} \, ,
\end{eqnarray}
%!p___________________________________________________________________
with the (unrenormalized) light-cone operators
%!p___________________________________________________________________
\begin{eqnarray}
\label{str_opra}
 O^\beta\kln{{\tx}/{2} ,-{\tx}/{2} }
&\equiv&
 \mbox{$\frac{\im}{2}$} \Klno{ : \bar\psi\kln{\tx/2} \,\gamma^\beta\,
\psi\kln{-\tx/2} : } \oKln{ \; - : \bar\psi\kln{-\tx/2} \,
\gamma^\beta\, \psi\kln{\tx/2} :} \, ,
\\
\label{str_oprb}
 O^{5 \, \beta}\kln{{\tx}/{2} ,-{\tx}/{2} }
&\equiv&
 \mbox{$\frac{\im}{2}$} \Klno{ : \bar\psi\kln{\tx/2}
\,\gamma^5\gamma^\beta\, \psi\kln{-\tx/2} : } \oKln{ \; + :
\bar\psi\kln{-\tx/2} \,\gamma^5\gamma^\beta \, \psi\kln{\tx/2} :} \, .
\end{eqnarray}
%!p___________________________________________________________________
This expansion has to be viewed as a simple form of the non-local
light-cone expansion (\ref{int_lce}) from which the suitable
$\Gamma$-structures can be read off. Considering general gauges also
the phase factors $U\kln{\kappa_1 \tx, \kappa_2 \tx}$ must be included
into the light-cone operators $O^\beta$ and $O^{5 \, \beta}$. Taking
into account also higher order terms of the $S$-matrix additional
operator structures come into the play which, of course, are
sub-leading.

The $\kappa$-integration appearing in the light-cone expansion
(\ref{int_lce}) connects the coefficient functions, given here at Born
level,
%!p___________________________________________________________________
\begin{eqnarray}
 && C_{1/2}'\kln{x^2,\kappa_1,\kappa_2} = \frac{\im }{2\pi^2 \kln{x^2
- \im \epsilon}^2} \; \KLe{ \delta\kln{\kappa_1 -
\hbox{$\frac{1}{2}$}} \delta\kln{\kappa_2 + \hbox{$\frac{1}{2}$}} \mp
\delta\kln{\kappa_2 - \hbox{$\frac{1}{2}$}} \delta\kln{\kappa_1 +
\hbox{$\frac{1}{2}$}} }
\end{eqnarray}
%!p___________________________________________________________________
with the respective operators. As is known from the general analysis
of the light-cone expansion \cite{MRGHD} the Fourier transforms of the
coefficient functions in Eq.~(\ref{int_lce}) are entire analytic
functions with respect to the variables $\tx p_i$ which leads to a
restriction of the integration range of $\kappa_1$ and $\kappa_2$ to
the interval $\kle{-1,+1}$.

For later convenience we introduce the variables
%!p___________________________________________________________________
\begin{equation}
 \kappa_\pm \equiv \hbox{$\frac{1}{2}$}\kln{\kappa_2 \pm \kappa_1}
\qquad \text{with} \qquad \kappa_{1,2} = \kappa_+ \mp \kappa_-
\end{equation}
%!p___________________________________________________________________
and use the following integration measure in the light-cone
expansion
%!p___________________________________________________________________
\begin{eqnarray}
\nonumber
 \D^2 \! \kappa
&\equiv&
 \! \d\kappa_1\, \d\kappa_2 \; \theta\kln{1-\kappa_1}
\theta\kln{1+\kappa_1} \theta\kln{1-\kappa_2} \theta\kln{1+\kappa_2}
\\
\nonumber
&=&
 \! 2 \, \d\kappa_+\, \d\kappa_- \; \theta\kln{1-\kappa_+ + \kappa_-}
\theta\kln{1+\kappa_+ - \kappa_-} \, \theta\kln{1-\kappa_+ - \kappa_-}
\theta\kln{1+\kappa_+ + \kappa_-} \; .
\end{eqnarray}
%!p___________________________________________________________________
In terms of the variables $\kappa_+$ and $\kappa_-$ the coefficient
functions read:
%!p___________________________________________________________________
\begin{eqnarray}
 && C_{1/2} \kln{x^2,\kappa_+, \kappa_-} = \frac{\im }{4\pi^2\kln{x^2
- \im \epsilon}^2} \; \delta\kln{\kappa_+} \KLe{\delta\kln{\kappa_- +
\hbox{$\frac{1}{2}$}} \mp \delta\kln{\kappa_- - \hbox{$\frac{1}{2}$}}
} .
\end{eqnarray}
%!p___________________________________________________________________
Using these conventions one obtains the following representation of
the operator $\widehat T_{\mu\nu}(x)$:
%!p___________________________________________________________________
\begin{eqnarray}
\label{str_ope}
&&
 R \, T\kle{J_\mu\kln{\frac{x}{2}}J_\nu\kln{-\frac{x}{2} }{\cal S} }
\approx \int_{\R^2} \D^2\!\kappa \,
\KLEo{C_1\kln{x^2,\kappa_+,\kappa_-} {S_{\mu\nu|}}^{\alpha\beta}
\,\tx_\alpha\, O_\beta\Kln{\kln{\kappa_+ - \kappa_-} \tx ,
\kln{\kappa_+ + \kappa_-} \tx } }
\\
\nonumber
&&
 \qquad \qquad \qquad \qquad \qquad \qquad \qquad \qquad \qquad \oKLE{
- \; \im \, C_2\kln{x^2,\kappa_+,\kappa_-}
{\varepsilon_{\mu\nu}}^{\alpha\beta} \, \tx_\alpha\, O^5_\beta
\Kln{\kln{\kappa_+ - \kappa_-} \tx , \kln{\kappa_+ + \kappa_-} \tx } }
\; .
\end{eqnarray}
%!p___________________________________________________________________
In general, the renormalized non-local operators
$O_\beta\kln{\kappa_1 \tx, \kappa_2 \tx}$ and
$O^5_\beta\kln{\kappa_1 \tx, \kappa_2 \tx}$ containing the phase
factors $U\kln{\kappa_1\tx,\kappa_2\tx}$ are given by
%!p___________________________________________________________________
\begin{eqnarray}
\label{str_oprab}
 \hspace*{-0.75cm} O_\beta\kln{\kappa_1\tx,\kappa_2\tx}
&=&
 \mbox{$\frac{\im}{2}$} R\,T \KLe{\kln{: \bar\psi\kln{\kappa_1\tx}
\,\gamma_\beta\,U\kln{\kappa_1\tx,\kappa_2\tx}\,\psi\kln{\kappa_2\tx}
: - : \bar\psi\kln{\kappa_2\tx}
\,\gamma_\beta\,U\kln{\kappa_1\tx,\kappa_2\tx} \,
\psi\kln{\kappa_1\tx} : } {\cal S} }
\\
\nonumber
&&
\\
\label{str_oprbb}
 \hspace*{-0.75cm} O^5_ \beta\kln{\kappa_1\tx,\kappa_2\tx}
&=&
 \mbox{$\frac{\im}{2}$} R\,T \KLe{\kln{: \bar\psi\kln{\kappa_1\tx}
\,\gamma^5 \gamma_\beta\,U\kln{\kappa_1\tx,\kappa_2\tx} \,
\psi\kln{\kappa_2\tx} : + : \bar\psi\kln{\kappa_2\tx} \,\gamma^5
\gamma_\beta\,U\kln{\kappa_1\tx,\kappa_2\tx} \, \psi\kln{\kappa_1\tx}
: } {\cal S}} .
\end{eqnarray}
%!p___________________________________________________________________
In the following we assume that the operators $O_\beta$ and
$O_\beta^5$ are renormalized quantities.

%%%%%%%%%%%%%%%%%%%%%%%%%%%%%%%%%%%%%%%%%%%%%%%%%%%%%%%%%%%%%%%%%%%%%%
\section{Twist decomposition and matrix elements}
\label{twi}
\setcounter{equation}{0}
%%%%%%%%%%%%%%%%%%%%%%%%%%%%%%%%%%%%%%%%%%%%%%%%%%%%%%%%%%%%%%%%%%%%%%

\noindent
The non-local operators $O_\beta\kln{\kappa_1 \tx, \kappa_2 \tx}$ and
$O^5_\beta\kln{\kappa_1 \tx, \kappa_2 \tx}$, Eqs.~(\ref{str_oprab}) and
(\ref{str_oprbb}), contain contributions of different twist. Here, the
notion of twist is used in its original form \cite{GT71} as
%!p___________________________________________________________________
\begin{eqnarray}
\label{twist}
 \text{(geometric) twist~}(\tau) &=& \text{scale dimension~}(d) -
\text{Lorentz spin~}(j) \ .
\end{eqnarray}
%!p___________________________________________________________________
The operators appearing in the expansion (\ref{str_ope}) of the
Compton amplitude have to be decomposed into their various twist
parts. On the light-cone they contain contributions of twist-2, 3 and
4
%!p___________________________________________________________________
\begin{eqnarray}
 O_\beta
&=&
 O_\beta^{\twz} + O_\beta^{\text{tw3}} + O_\beta^{\text{tw4}}\ ,
\\
 O_\beta^5
&=&
 O_\beta^{5 \, \twz} + O_\beta^{5 \, \text{tw3}} + O_\beta^{5 \,
\text{tw4}}~;
\end{eqnarray}
%!p___________________________________________________________________
however, off the light-cone they contain an infinite series of
growing twist.

A group theoretical procedure of the twist decomposition has been
worked out in Refs.~\cite{GLR99,GL00a} and applied to various
physically relevant light-ray operators. As a result, the twist-2 part
of the operators (\ref{str_oprab}) and (\ref{str_oprbb}) can be
constructed out of the twist-2 (pseudo) scalar operators
$O^{(5)\twz}\kln{\kappa_1 \tx, \kappa_2 \tx} =
\tx^\beta O_\beta^{(5)}\kln{\kappa_1 \tx, \kappa_2 \tx}$
by applying the interior derivative
%!p___________________________________________________________________
\begin{equation}
 \text{d}_\beta = \kln{1 + \tx \tilde \partial} \, \tilde \partial_\beta -
\frac{1}{2} \tx_\beta \, \tilde\partial^2 \qquad{\rm with}\qquad
\text{d}^2 =0\ ,
\end{equation}
%!p___________________________________________________________________
on the un-decomposed light-cone operators and performing a
subsequent $\tau-$integration (which stems from the normalization of
the local operators):
%!p___________________________________________________________________
\begin{eqnarray}
\label{TW2}
 O_\beta^{\twz}\kln{\kappa_1 \tx, \kappa_2 \tx}
&=&
 - \int_0^1 \d\tau \; \ln\tau\; \text{d}_\beta \, O\kln{\kappa_1 \tau
\tx, \kappa_2 \tau \tx}
\\
\label{twi_tw2}
\nonumber
&=&
 \int_0^1 \d\tau \kle{\tilde\partial_\beta +\frac{1}{2}\ln\tau\;\tx_\beta
\; \tilde\partial^2} O \kln{\kappa_1 \tau \tx, \kappa_2 \tau \tx}\ .
\end{eqnarray}
%!p___________________________________________________________________
According to its structure the tracelessness of the operator
(\ref{TW2}) which corresponds on the light-cone to the requirement
$\d^\beta O_\beta^{\twz}\kln{\kappa_1 \tx, \kappa_2 \tx} = 0$, is
trivially fulfilled due to the property of $\d^\beta$. An analogous
relation holds for the axial vector and pseudo scalar operator. The
twist-3 and twist-4 parts, however, cannot be constructed out of the
(pseudo) scalar operator since the latter, when restricted to the
light-cone, is already of twist-2.

Since we want to extract the twist-2 part of the Compton amplitude,
relation (\ref{TW2}) will be applied to the matrix elements of the
operators considered~:
%!p___________________________________________________________________
\begin{eqnarray}
\label{twi_mat}
 \Matel{p_2,k}{O_\beta^{\twz}\kln{\kappa_1 \tx, \kappa_2 \tx}}{p_1} &=&
\int_0^1 \! \d\tau \! \kle{ \tilde\partial_\beta +\frac{1}{2} \ln\tau \,
\tx_\beta \, \tilde\partial^2} \! \Matel{p_2,k}{O\kln{\kappa_1 \tau \tx,
\kappa_2 \tau \tx}}{p_1} .
\end{eqnarray}
%!p___________________________________________________________________
Here, the spins of the nucleons have been suppressed in the
notation. Let us mention that the geometric twist decomposition of the
matrix elements is due to the twist decomposition of the non-local
operators. Usually, in phenomenological considerations another notion
of twist, called `dynamical' twist, is considered which has been
introduced in the decomposition of the (forward) matrix elements by
Jaffe and Ji \cite{JJ91}. The interrelation of geometric and dynamic
twist was considered in Ref.~\cite{GL01}. In the case of lowest
twist-2 there appears no difference, but for higher twist the mismatch
of dynamical twist with respect to geometric twist leads to differing
structures.

Because of translation invariance,
%!p___________________________________________________________________
\begin{eqnarray}
 \Matel{p_2,k}{O\kln{\kappa_1 \tx, \kappa_2 \tx}}{p_1} &=& \e^{\im
\kappa_+ \, \tx P_-} \Matel{p_2,k}{O\kln{-\kappa_- \tx, \kappa_-
\tx}}{p_1} \; ,
\end{eqnarray}
%!p___________________________________________________________________
it is more convenient to discuss the centered operator
$O\kln{-\kappa \tx, \kappa \tx}$. Henceforth, for brevity, $\kappa_-$
will be denoted by $\kappa$.

In the ordinary non-forward case one usually parameterizes the
scalar matrix element by a Dirac- and a Pauli-type
contribution,
%!p___________________________________________________________________
\begin{eqnarray}
 \KI_1\kln{\tx,p_2,p_1}
&=&
 \tx^\beta \; \bar u\kln{p_2} \, \gamma_\beta \, u\kln{p_1}\ ,
\\
 \KI_2\kln{\tx,p_2,p_1}
&=&
 \tx^\beta \; \frac{1}{m_0} \bar u\kln{p_2} \, \sigma_{\beta\alpha} p_-^\alpha
\, u\kln{p_1} \ ,
\end{eqnarray}
%!p___________________________________________________________________
where $u\kln{p_1}$ and $\bar u\kln{p_2}$ are on-shell spinors of the
incoming and outgoing nucleons, $m_0$ is a dimensional mass scale
which is kept fixed, and
$\sigma_{\beta\alpha} = (\im /2) \kom{\gamma_\beta}{\gamma_\alpha}$. Because of
$p_- \ra 0$ the Pauli type factor ${\KI}_{2}$ vanishes in the forward
limit. Using these kinematic factors the matrix element can be
parameterized as follows:
%!p___________________________________________________________________
\begin{eqnarray}
\label{twi_sca}
 \Matel{p_2}{O\kln{-\kappa \tx, \kappa \tx}}{p_1} &=& \sum_{a=1}^2
\KI_a\kln{\tx,p_2,p_1}\, \tilde f_a\kln{\kappa \tx p_+, \,\kappa \tx
p_-, \,p_i p_j, \, \mu^2} \ .
\end{eqnarray}
%!p___________________________________________________________________
Here, the coefficient functions $\tilde f_a$ are the distribution
amplitudes in $x-$space. They depend on $\kappa \tx p_i$ and all
possible products of the two momenta,
$p_i p_j = \kls{p_1^2, p_2^2, p_1 p_2}$, as well as on the
renormalization scale $\mu$ and the coupling constant $g$.

If an outgoing meson is present in the process, which means that
one has to contruct the matrix elements
$\Matel{p_2,k}{O(-\kappa \tx, \kappa \tx)}{p_1}$, the representation
(\ref{twi_sca}) of the scalar matrix element must be modified because
of the presence of the additional momentum $k$. Especially, further
kinematic structures occur. They can be determined in a
straightforward way by using the following parameterization of the
scalar matrix element:
%!p___________________________________________________________________
\begin{equation}
\label{scalar_mat}
 \Matel{p_2,k}{O\kln{-\kappa \tx, \kappa \tx}}{p_1} = \bar u\kln{p_2}
\, \Lambda\kln{\kappa \tx, p_2, k, p_1} \, u\kln{p_1},
\end{equation}
%!p___________________________________________________________________
with
%!p___________________________________________________________________
\begin{equation}
 \Lambda\kln{\kappa \tx, p_2, k, p_1} = \mathbf{ A} \, \Id + \mathbf{
B}_\alpha\, \gamma^\alpha + \mathbf{ C}_{\kle{\alpha\beta}} \,
\sigma^{\alpha\beta} + \mathbf{ D}_\alpha \, \gamma^5\gamma^\alpha \;
.
\end{equation}
%!p___________________________________________________________________
It is an easy task to find the general structure of
$\mathbf{A}-\mathbf{D}$ allowed by the momenta
$p_{1\alpha}, p_{2\alpha}, k_\alpha$, the metric $g_{\alpha\beta}$ and the
Levi-Civita tensor $\varepsilon_{\beta\alpha\gamma\delta}$, demanding
$\Lambda$ not to be a pseudo scalar.

Using the equations of motion for the hadronic momenta $p_{1\mu}$
and $p_{2\mu}$ with $M$ denoting the nucleon mass,
%!p___________________________________________________________________
\begin{eqnarray}
 \gamma^\alpha p_{1\alpha} \; u\kln{p_1}
&=&
 M \, u\kln{p_1}\ ,
\\
 \bar u\kln{p_2} \, \gamma^\alpha p_{2\alpha}
&=&
 M \, \bar u\kln{p_2} \ ,
\end{eqnarray}
%!p___________________________________________________________________
the decomposition of the matrix element (\ref{scalar_mat}) is~:
%!p___________________________________________________________________
\begin{eqnarray}
 \KI'_1
&=&
 (\bar u \, u)\ ,
\\
\nonumber
 \KI'_2
&=&
 (\bar u\,\Fe{\tx} \, u)\ ,
\\
\nonumber
 \KI'_3
&=&
 \frac{1}{m_0} \; (\bar u\,\Fe{k} \, u)\ ,
\\
\nonumber
 \KI'_4
&=&
 \frac{1}{m_0} \; (\bar u\,\Fe{\tx} \Fe{k}\, u)\ ,
\\
\nonumber
 \KI'_5
&=&
 \frac{1}{m_0^3} \; \im \,\varepsilon_{\beta\alpha\gamma\delta} \; \tx^\beta
\, p_2^\alpha \, k^\gamma \, p_1^\delta\; (\bar u\; \gamma^5\; u) \ ,
\end{eqnarray}
%!p___________________________________________________________________
where an auxiliary mass $m_0$ has been introduced in order to get
kinematic structures of equal dimensionality. Then, the scalar matrix
element is parameterized by a sum over the five kinematic factors
$\KI_a'$ as follows:
%!p___________________________________________________________________
\begin{eqnarray}
\label{twi_zer}
 \Matel{p_2,k}{O\kln{-\kappa \tx, \kappa \tx}}{p_1} &=& \sum_{a=1}^5
\KI'_a\kln{\tx,{\mathbf{p}}} \, \tilde f'_a\kln{\kappa \tx
{\mathbf{p}}, \, {\mathbf{p}}_i {\mathbf{p}}_j ,\, \mu^2 } \ ,
\end{eqnarray}
%!p___________________________________________________________________
where ${\mathbf{p}}\equiv \kls{{\mathbf p}_i} = \kls{p_1, p_2, k}$
generically denotes the multi-vector in the space of all the (three)
hadronic momenta.

Although one possible set of kinematic factors is given by
(\ref{twi_zer}), it will be more convenient to choose another one
which is also linearly independent and contains the original kinematic
factors $\KI_a'$. Mimicking the Dirac- and Pauli-structures we choose
%!p___________________________________________________________________
\begin{eqnarray}
 \KI_1\kln{\tx,p_2,k,p_1}
&=&
 \tx^\beta\, (\bar u \, \gamma_\beta\, u)\ ,
\\
\nonumber
 \KI_2\kln{\tx,p_2,k,p_1}
&=&
 \tx^\beta\, \frac{1}{m_0} (\bar u\, \sigma_{\beta\alpha} \, P_-^\alpha\, u)\ ,
\\
\nonumber
 \KI_3\kln{\tx,p_2,k,p_1}
&=&
 \tx^\beta \, \frac{1}{m_0^2} (\bar u\, k_\beta \, \gamma_{\alpha}\, k^\alpha \,
u) \ ,
\\
\nonumber
 \KI_4\kln{\tx,p_2,k,p_1}
&=&
 \tx^\beta \, \frac{1}{m_0} (\bar u\, \sigma_{\beta\alpha} k^\alpha \, u)\ ,
\\
\nonumber
 \KI_5\kln{\tx,p_2,k,p_1}
&=&
 \tx^\beta \; \frac{1}{m_0^3} \im \,\varepsilon_{\beta\alpha\gamma\delta} \,
p_2^\alpha \, k^\gamma \, p_1^\delta \, (\bar u \,\gamma^5\, u)\ ,
\end{eqnarray}
%!p___________________________________________________________________
for the matrix element of the scalar operator and
%!p___________________________________________________________________
\begin{eqnarray}
 \KI_1^5\kln{\tx,p_2,k,p_1}
&=&
 \tx^\beta\,(\bar u \,\gamma^5 \,\gamma_\beta \, u)\ ,
\\
\nonumber
 \KI_2^5\kln{\tx,p_2,k,p_1}
&=&
 \tx^\beta\, \frac{1}{m_0} (\bar u\, \gamma^5 \, \sigma_{\beta\alpha}\,
P_-^\alpha\, u)\ ,
\\
\nonumber
 \KI_3^5\kln{\tx,p_2,k,p_1}
&=&
 \tx^\beta \, \frac{1}{m_0^2} (\bar u\, \gamma^5 \, k_\beta \gamma_{\alpha}
\, k^\alpha \, u)\ ,
\\
\nonumber
 \KI_4^5\kln{\tx,p_2,k,p_1}
&=&
 \tx^\beta \, \frac{1}{m_0}(\bar u \, \gamma^5 \, \sigma_{\beta\alpha} k^\alpha
\, u)\ ,
\\
\nonumber
 \KI_5^5\kln{\tx,p_2,k,p_1}
&=&
 \tx^\beta \; \frac{1}{m_0^3} \im \,\varepsilon_{\beta\alpha\gamma\delta} \,
p_2^\alpha \, k^\gamma \, p_1^\delta (\bar u\; u)\ ,
\end{eqnarray}
%!p___________________________________________________________________
for the matrix element of the pseudo scalar operator. For explicit
calculations it is important to note that each factor $\KI_a^{(5)}$
has a linear $\tx$-dependence,
%!p___________________________________________________________________
\begin{eqnarray*}
 \KI_a(\tx,{\mathbf p})
&=&
 \tx^\beta \, \KI_{a \beta}({\mathbf p})\ ,
\\
 \KI_a^5(\tx,{\mathbf p})
&=&
 \tx^\beta \, \KI^5_{a \beta}({\mathbf p}) \ .
\end{eqnarray*}
%!p___________________________________________________________________
Using the equations of motion again, the factors $\KI_a$ equal the
following combinations of $\KI_a'$
%!p___________________________________________________________________
\begin{eqnarray}
 \KI_1
&=&
 \KI'_2\ ,
\\
\nonumber
 \KI_2
&=&
 \frac{\im}{m_0} \kln{\tx p_2 + \tx k + \tx p_1} \; \KI'_1 - 2\im \,
\frac{M}{m_0} \; \KI'_2 - \im \, \KI'_4\ ,
\\
\nonumber
 \KI_3
&=&
 \frac{1}{m_0} (\tx k) \; \KI'_3\ ,
\\
\nonumber
 \KI_4
&=&
 \frac{\im}{m_0} \, (\tx k) \; \KI'_1 - \im \, \KI'_4 \ ,
\\
\nonumber
 \KI_5
&=&
 \KI'_5 \, .
\end{eqnarray}
%!p___________________________________________________________________
This shows that the $\KI_a$ constitute a suitable set of kinematic
factors for the scalar matrix element. In the limit $k \ra 0$ they
obviously reduce to the Dirac- and Pauli-structures. Analogous
statements hold for $\KI^5_a$.

The decomposition of the matrix element of
$O\kln{-\kappa \tx, \kappa \tx}$ and
$O^5\kln{-\kappa \tx, \kappa \tx}$ now reads
%!p___________________________________________________________________
\begin{eqnarray}
\label{mat_zerl}
 \Matel{p_2,k}{O\kln{-\kappa \tx, \kappa \tx}}{p_1}
&=&
 \sum_{a=1}^5 \KI_a \kln{\tx,{\mathbf p}} \tilde f_a\kln{\kappa \tx
{\mathbf p}, \, {\mathbf{p}}_i {\mathbf{p}}_j , \, \mu^2 }\ ,
\\
 \Matel{p_2,k}{O^5\kln{-\kappa \tx, \kappa \tx}}{p_1}
&=&
 \sum_{a=1}^5 \KI_a^5 \kln{\tx,{\mathbf p}} \tilde f_a^5\kln{\kappa
\tx {\mathbf p}, \, {\mathbf{p}}_i {\mathbf{p}}_j , \, \mu^2 } \, .
\end{eqnarray}
%!p___________________________________________________________________
In principle there is also a dependence on variables like
${\mathbf{p}}_i {\mathbf{p}}_j \, x^2$. Because the latter dependence
vanishes on the light-cone, it will not be discussed in the further
considerations. As has been shown in Ref.~\cite{GLR01} the whole
$x^2-$dependence is governed by harmonic extension off the light-cone
if the operator structure is already given on--cone. The
$\mu^2-$dependence is governed by the evolution equations obeyed by
the functions $\tilde f_a$ and will be discussed in Section~\ref{evo}.

As next step, a Fourier transformation of the functions $\tilde f_a$
is performed,
%!p___________________________________________________________________
\begin{equation}
 \tilde f_a \kln{\kappa \tx {\mathbf p}} = \int_{\R^3} \! \D^3 \! z \; \;
\e^{-\im \kappa \tx\kln{{\mathbf p}{\mathbf z} } } f_a\kln{{\mathbf
z}}\ .
\end{equation}
%!p___________________________________________________________________
Because also the functions $\tilde f_a\kln{\kappa \tx {\mathbf p}}$
are entire analytic in the variables $\kappa \tx {\mathbf p}$, the
support of their Fourier transforms $f_a\kln{{\mathbf z}}$ is
restricted to the interval $\kle{-1,+1}$ in the variables
${\mathbf z}=\kln{z_1,z_2,z_3}$. Therefore, the measure
%!p___________________________________________________________________
\begin{eqnarray}
\label{ddreiz}
 \D^3 \! z &\equiv& \d z_1 \, \d z_2 \, \d z_3 \; \theta\kln{1-z_1}
\theta\kln{1+z_1} \theta\kln{1-z_2} \, \theta\kln{1+z_2}
\theta\kln{1-z_3} \theta\kln{1+z_3}
\end{eqnarray}
%!p___________________________________________________________________
has been introduced to realize this support.
${\mathbf p}{\mathbf z}$ is simply the product of the vectors
${\mathbf p}$ and ${\mathbf z}$, see (\ref{PU}). To get a
representation in the momenta $P_\pm$ and $k$ one introduces the
variables
%!p___________________________________________________________________
\begin{equation*}
 z_+ = \frac{1}{2}\kln{z_1 + z_2}, \quad z_- = \frac{1}{2}\kln{ z_2 -
z_1}, \quad z_k = z_3 - z_2,
\end{equation*}
%!p___________________________________________________________________
with
%!p___________________________________________________________________
\begin{equation*}
 z_1 = z_+ - z_-, \quad z_2 = z_+ + z_- , \quad z_3 = z_+ + z_- + z_k
\ .
\end{equation*}
%!p___________________________________________________________________
Using the abbreviation
%!p___________________________________________________________________
\begin{eqnarray}
\label{PU}
 \PU
&\equiv&
 P_+ z_+ + P_- z_- + k \, z_k
\\
\nonumber
&=&
 p_1 \, z_1 + p_2 \, z_2 + k \, z_3 \; \equiv \; {\mathbf p}{\mathbf
z} \, ,
\end{eqnarray}
%!p___________________________________________________________________
the scalar matrix element is represented by
%!p___________________________________________________________________
\begin{eqnarray}
\label{scalar_matrix_element}
 \Matel{p_2,k}{O\kln{-\kappa \tx, \kappa \tx}}{p_1} &=& \sum_{a=1}^5
\KI^a\kln{\tx,{\mathbf p}} \int_{\R^3} \! \D^3 \! z \;\; \e^{-\im \kappa
\, \tx\PU} f_a\kln{z_+,z_-,z_k} \, .
\end{eqnarray}
%!p___________________________________________________________________
In the following the explicit summation over $a$ will be omitted,
but will be indicated by the position of the index $a$.

The expression (\ref{scalar_matrix_element}) will be inserted into
(\ref{twi_mat}) to calculate the matrix element of $O_\mu^{\twz}$.
Thereby, it is important to state that the $\tau-$scaling in equation
(\ref{twi_mat}) refers to $\kappa$ and not to $\tx$. Performing a
change of variables $z_\pm \ra z_\pm / \tau$ and $z_k \ra z_k / \tau$
we get the following form of the matrix element, for the general case
$\kappa_1 \neq - \kappa_2$:
%!p___________________________________________________________________
\begin{eqnarray}
 \matel{p_2,k}{O_\beta^{\twz}\kln{\kappa_1 \tx, \kappa_2 \tx}}{p_1}
&=&
 \int_{\R^3} \d z_+ \, \d z_- \, \d z_k \; \int_0^1 \d\tau \,
\kle{\tilde\partial_\beta + \frac{1}{2} \ln\tau \;\, \tx_\beta \,
\tilde\partial^2} \; \frac{1}{\tau^3}
\\
\nonumber
&&
 \quad \times \; \e^{\im \tx\kln{\kappa_+ \tau \, P_- - \kappa \, \PU}
} \; \tx_\rho \; \KI^{a\rho} \kln{{\mathbf p}} \;
f_a\kln{\frac{z_+}{\tau},\frac{z_-}{\tau},\frac{z_k}{\tau}}
\Theta\kln{\tau,z_+,z_-,z_k} \ ,
\end{eqnarray}
%!p___________________________________________________________________
where the abbreviation
%!p___________________________________________________________________
\begin{eqnarray}
\nonumber
 \Theta\kln{\tau,z_+,z_-,z_k}
&\equiv&
 2\, \theta\kln{\tau-z_+ + z_- }\theta\kln{\tau+z_+ - z_-}
\theta\kln{\tau-z_+ - z_- } \theta\kln{\tau+z_+ + z_- }
\\
&&
 \; \times \theta\kln{\tau-z_+ - z_- - z_k } \theta\kln{\tau+z_+ + z_-
+ z_k }
\end{eqnarray}
%!p___________________________________________________________________
has been used. Carrying out the differentiations we get
%!p___________________________________________________________________
\begin{eqnarray}
\label{unsymmetrisch}
\nonumber
&&
 \matel{p_2,k}{O_\beta^{\twz}\kln{\kappa_1 \tx, \kappa_2 \tx}}{p_1} =
\int_{\R^3}\!\! \d z_+ \, \d z_- \, \d z_k \!\! \int_0^1
\frac{\d\tau}{\tau^3} \; \e^{\im \tx\kln{\kappa_+ \tau \, P_- - \kappa
\, \PU} } \; \KI^{a \rho} ({\mathbf p}) \,
f_a\!\kln{\frac{z_+}{\tau},\frac{z_-}{\tau},\frac{z_k}{\tau}} \;
\Theta\kln{\tau,z_+,z_-,z_k}
\\
\nonumber
&&
 \qquad \qquad \qquad \qquad \qquad \times \, \KLE{g_{\beta\rho} +
\im\kln{\kappa_+\tau P_{-\beta}-\kappa \PU_\beta } \tx_\rho -
\frac{\tx_\beta}{2}\,\ln\tau \kln{\kln{\kappa_+ \tau P_- - \kappa \,\PU
}^2 \tx_\rho - 2\im \kln{ \kappa_+ \tau P_{-\rho} - \kappa \, \PU_\rho
}}}
\\
&&
\end{eqnarray}
%!p___________________________________________________________________
This form of the matrix element is yet rather complicated. Since
only the centered operator is needed in the following considerations,
we set $\kappa_+ = 0$.

To derive a simpler representation for the matrix element, the
$\tau$-integration will now be comprised into the functions $F_a$ and
$F_a^{\text{tr}}$, the latter denoting the trace part, defined by
%!p___________________________________________________________________
\begin{eqnarray}
\label{Fa}
 F_a({\mathbf z})
&\equiv&
 F_a \kln{z_+,z_-,z_k} =\int_0^1 \d\tau \; \frac{1}{\tau^3} \; \hat
f_a\kln{\frac{z_+}{\tau},\frac{z_-}{\tau},\frac{z_k}{\tau}}\ ,
\\
\label{Ga}
 F_a^{\mathrm{tr}}({\mathbf z})
&\equiv&
 F_a^{\mathrm{tr}} \kln{z_+,z_-,z_k} = \int_0^1 \d\tau \;
\frac{\ln\tau}{\tau^3} \; \hat f_a\kln{\frac{z_+}{\tau},
\frac{z_-}{\tau},\frac{z_k}{\tau}} \ ,
\end{eqnarray}
%!p___________________________________________________________________
with $\hat f_a$ given by
%!p___________________________________________________________________
\begin{equation}
\label{hat_fa}
 \hat f_a\kln{\frac{z_+}{\tau},\frac{z_-}{\tau},\frac{z_k}{\tau}}
\equiv f_a\kln{\frac{z_+}{\tau},\frac{z_-}{\tau},\frac{z_k}{\tau}} \;
\Theta\kln{\tau,z_+,z_-,z_k}~.
\end{equation}
%!p___________________________________________________________________
Obviously, these distribution amplitudes are not independent, and
the restricted integration range in $z-$space finally is contained in
the support properties of the distribution amplitudes
$F_a\kln{z_+,z_-,z_k}$ and $F_a^{\text{tr}} \kln{z_+,z_-,z_k}$.

After the substitution of $F_a$ and $F_a^{\text{tr}}$ in
(\ref{unsymmetrisch}) with $\kappa_+ = 0$ one obtains
%!p___________________________________________________________________
\begin{eqnarray}
\label{twi_enda}
&&
 \matel{p_2,k}{O_\beta^{\twz}\kln{-\kappa \tx, \kappa \tx}}{p_1}
\\
\nonumber
&&
 \qquad = \int_{\R^3} \d z_+ \, \d z_- \, \d z_k \;\; \e^{ - \im
\kappa \, \tx\PU }\, \KLEo{\Kln{ g_{\beta\rho} - \im \kappa \,\PU_\beta
\tx_\rho} F_a({\mathbf z}) } \oKLE{ - \frac{\tx_\beta}{2}
\kln{\kappa^2\PU^2 \tx_\rho + 2\im \kappa \, \PU_\rho }
F_a^{\text{tr}}({\mathbf z})} \, \KI^{a \rho}({\mathbf p})
\end{eqnarray}
%!p___________________________________________________________________
for the matrix element of the twist-2 part of the vector operator
$O_\mu$. The matrix element of the axial vector operator
$O_\mu^{5 \, \twz}$ possesses a similar representation
%!p___________________________________________________________________
\begin{eqnarray}
\label{twi_endb}
&&
 \matel{p_2,k}{O_\beta^{5 \, \twz} \kln{-\kappa \tx, \kappa \tx}}{p_1}
\\
\nonumber
&&
 \qquad = \int_{\R^3} \d z_+ \, \d z_- \, \d z_k \;\; \e^{ - \im
\kappa \, \tx\PU }\, \KLEo{\Kln{ g_{\beta\rho} - \im \kappa \,\PU_\beta
\tx_\rho} F_a^5({\mathbf z}) } \oKLE{ - \frac{\tx_\beta}{2}
\kln{\kappa^2\PU^2 \tx_\rho + 2\im \kappa \, \PU_\rho }
F_a^{5\,\text{tr}}({\mathbf z})} \, \KI^{5 \,a\rho} ({\mathbf p})~,
\end{eqnarray}
%!p___________________________________________________________________
with $F_a^5$ and $F_a^{5\,\text{tr}} $ defined analogously to $F_a$
and $F_a^{\text{tr}}$, by exchanging $\hat f_a$ and $\hat f_a^5$ in
(\ref{Fa}) and (\ref{Ga}).

Here, some general remarks are in order. First, the {\em
triple-valued distribution amplitudes} $F_a^{(5)}\kln{z_+,z_-,z_k}$
and $F_a^{(5)\,\text{tr}}\kln{z_+,z_-,z_k}$ are uniquely related to
the twist-2 (axial) vector operators and, in principle, should have
been marked by the related twist $\tau = 2$. However, since we do not
consider higher twist this has been omitted. Second, every
distribution amplitude of definite twist -- also off the light-cone --
depends only on the ${\mathbf z}-$variables and, possibly, on the
momenta $\mathbf p$. The $x-$dependence is completely contained in the
accompanying factors including, of course, the exponential
$\e^{ - \im \kappa \, x\PU }$. These general expressions which are
given in terms of Bessel functions of the argument
$(\kappa/2)\sqrt{(x\PU)^2 - x^2 \PU^2}$ have been determined in
Ref.~\cite{GLR01}
(For the twist--2 case of DIS this statement has already
been made in Ref.~\cite{BB91}.)
Restricting onto the light-cone leads to the
expressions (\ref{twi_enda}) and (\ref{twi_endb}). 
Third, these properties
hold for the operators of definite twist and are transposed to the
corresponding matrix elements, independently how many particles
(momenta) occur in the incoming and outgoing states. Therefore, if a
Fourier transformation containing these matrix elements has to be
performed this can be done by simply replacing $\tx \ra x$. (In
Appendix A we give these expressions explicitly together with their
restriction onto the light-cone.) This will be applied in the next
Section for the expressions Eqs.~(\ref{twi_enda}) and
(\ref{twi_endb}).

%%%%%%%%%%%%%%%%%%%%%%%%%%%%%%%%%%%%%%%%%%%%%%%%%%%%%%%%%%%%%%%%%%%%%%
\section{Compton amplitude}
\label{com}
\setcounter{equation}{0}
%%%%%%%%%%%%%%%%%%%%%%%%%%%%%%%%%%%%%%%%%%%%%%%%%%%%%%%%%%%%%%%%%%%%%%

\noindent
Now, we are in a position to compute the twist-2 part of the Compton
amplitude (\ref{Compton_amp}). Of course, we need it in the extended
Bjorken region and, therefore, can restrict our consideration to the
neighborhood of the light-cone. Thus, we will take the matrix elements
(\ref{twi_enda}) and (\ref{twi_endb}) with $\tx$ replaced by $x$.
Merging everything together, we use the representation (\ref{str_ope})
for the time-ordered product of two hadronic currents, insert
(\ref{twi_enda}) and (\ref{twi_endb}) for the matrix elements of
$O_\beta^{\twz}$ and $O_\beta^{5 \, \twz}$ and obtain:
%!p___________________________________________________________________
\begin{eqnarray}
 T_{\mu\nu}^{\twz}\kln{p_2,k,p_1,q}
&=&
 \im \ix \; \e^{\im qx} \int_{\R^2}\!\! \D^2\!\kappa \; x_\alpha
\KLEo{C_1\kln{x^2,\kappa_+,\kappa} {S_{\mu\nu|}}^{\alpha\beta}
\MAtel{p_2,k}{O^\twz_\beta\Kln{\kln{\kappa_+ - \kappa} x ,
\kln{\kappa_+ + \kappa} x}}{p_1} }
\\
\nonumber
&&
 \qquad\qquad\qquad\qquad\qquad\quad \oKLE{- \im \,
C_2\kln{x^2,\kappa_+,\kappa} {\varepsilon_{\mu\nu}}^{\alpha\beta}
\MAtel{p_2,k}{O_\beta^{5\,\twz} \Kln{\kln{\kappa_+ - \kappa} x ,
\kln{\kappa_+ + \kappa} x }}{p_1} }\ .
\end{eqnarray}
%!p___________________________________________________________________
The arguments $\kappa_+$ and $\kappa$ of the coefficient functions
$C_i\kln{x^2,\kappa_+,\kappa}$ are fixed at $\kappa_+=0$ and
$\kappa =\pm 1/2$, so that we can use the symmetry properties of the
operators $O_\beta$ and $O^5_\beta$,
%!p___________________________________________________________________
\begin{eqnarray}
 O_\beta\kln{-\kappa x,\kappa x}
&=&
 -O_\beta\kln{\kappa x, -\kappa x}\ ,
\\
\nonumber
 O^5_\beta\kln{-\kappa x,\kappa x}
&=&
 O_\beta^5\kln{\kappa x,-\kappa x}\ .
\end{eqnarray}
%!p___________________________________________________________________
For the computation of the Compton amplitude it suffices to know the
operators at these given points, but for the investigation of the
evolution of the matrix elements their representation at general
values of $\kappa_+$ and $\kappa$ is needed, see Section~\ref{evo} for
the details.

Performing the $\kappa$-integration one obtains
%!p___________________________________________________________________
\begin{eqnarray}
&&
 T_{\mu\nu}^{\twz}\kln{p_2,k,p_1,q} = - 2 \int_{\R^3} \d z_+ \, \d z_-
\, \d z_k \int \frac{\d^4 \! x}{2\pi^2} \; \e^{\im\QU x}
\frac{x_\alpha} { \kln{x^2 - \im \epsilon}^2 }
\\
\nonumber
&&
 \qquad\quad \times \ \kls{{ S_{\mu\nu|}}^{\alpha\beta} \kle{\kln{
g_{\beta\rho} - \frac{\im}{2} \PU_\beta \; x_\rho } F_a - x_\beta
\kln{\frac{1 }{8}\PU^2 \; x_\rho + \frac{\im}{2} \PU_\rho }
F_a^{\text{tr}} } \; \KI^{a\rho} - \im
{\varepsilon_{\mu\nu}}^{\alpha\beta} \kln{ g_{\beta\rho} -
\frac{\im}{2} \PU_\beta \; x_\rho } \, F_a^5 \; \KI^{5\,a\rho} } \ ,
\end{eqnarray}
%!p___________________________________________________________________
where the abbreviation
%!p___________________________________________________________________
\begin{equation}
 \QU \equiv q - \frac{1}{2} \PU = q - \frac{1}{2} \kln{P_+ z_+ + P_-
z_- + k \, z_k}
\end{equation}
%!p___________________________________________________________________
has been used. The $x_\beta$-term which results from the traces of
the twist-2 operator vanishes for the axial matrix element because
$x_\alpha x_\beta$ is symmetric. As a last step in the computation of
the Compton amplitude the Fourier transformation is carried out by
using
%!p___________________________________________________________________
\begin{eqnarray*}
 \int \frac{\d^4 \! x}{2\pi^2} \; \e^{\im\QU x}
\frac{x_\alpha}{\kln{x^2 - \im \epsilon}^2}
&=&
 \frac{\QU_\alpha}{\QU^2 + \im \epsilon} \, ,
\\
 \int \frac{\d^4 \! x}{2\pi^2} \; \e^{\im\QU x} \frac{x_\alpha
x_\beta}{\kln{x^2 - \im \epsilon}^2}
&=&
 - \im \frac{g_{\alpha\beta}}{\QU^2 + \im \epsilon} + 2 \im
\frac{\QU_\alpha \QU_\beta}{\kln{\QU^2 + \im \epsilon}^2} \, ,
\\
 \int \frac{\d^4 \! x}{2\pi^2} \; \e^{\im\QU x} \frac{x_\alpha x_\beta
x_\rho}{\kln{x^2 - \im \epsilon}^2}
&=&
 2\frac{g_{\alpha\beta}\QU_\rho + g_{\beta\rho}\QU_\alpha +
g_{\alpha\rho} \QU_\beta}{ \kln{\QU^2 + \im \epsilon}^2 } - 8
\frac{\QU_\alpha \QU_\beta \QU_\rho} {\kln{\QU^2 + \im \epsilon}^3 }
\, ,
\end{eqnarray*}
%!p___________________________________________________________________
and the summation over $\alpha$ and $\beta$ is performed in the
symmetric part of $T_{\mu\nu}^\twz$ by using the form (\ref{SYM}) for
${S_{\mu\nu|}}^{\alpha\beta}$. Then for the Compton amplitude we get
the result
%!p___________________________________________________________________
\begin{eqnarray}
\label{Compton_b}
 T_{\mu\nu}^\twz
&=&
 - 2 \int_{\R^3} \d z_+ \, \d z_- \, \d z_k \; \; \frac{1}{\QU^2 + \im
\epsilon} \KLEEo{ - \im \, {\varepsilon_{\mu\nu}}^{\alpha\beta} \;\,
q_\alpha \kls{ g_{\beta\rho} + \frac{ \PU_\beta \, \QU_\rho}{\QU^2 +
\im \epsilon} } \; F_a^5 \; \KI^{5\,a\rho} }
\\
\nonumber
&&
 + \KLS{g_{\mu\rho}\kln{q_\nu - \PU_\nu} + g_{\nu\rho}\kln{q_\mu -
\PU_\mu} - g_{\mu\nu}\kln{q_\rho - \PU_\rho} +\frac{\QU_\rho}{\QU^2 +
\im \epsilon} \KLe{ \QU_\mu \PU_\nu + \QU_\nu \PU_\mu - g_{\mu\nu} \,
\PU\QU }} \; F_a \; \KI^{a\rho}
\\
\nonumber
&&
\\
\nonumber
&&
 + \KLSo{ g_{\mu\nu}\PU_\rho + \frac{1}{\QU^2 + \im \epsilon}
\kle{\KLn{ 2 \, \QU_\mu \QU_\nu - g_{\mu\nu} \QU^2 } \PU_\rho -
\frac{1}{2}\KLn{ g_{\nu\rho}\QU_\mu + g_{\mu\rho} \QU_\nu - 2 \,
g_{\mu\nu}\QU_\rho } \PU^2 } }
\\
\nonumber
&&
 \oKLEE{ \qquad \qquad \oKLS{ + \frac{\QU_\rho \, \PU^2}{ \kln{\QU^2 +
\im \epsilon}^2} \KLn{2 \, \QU_\mu \QU_\nu - g_{\mu\nu} \QU^2 } } \;
F_a^{\text{tr}} \; \KI^{a\rho} \;\; } \ ,
\end{eqnarray}
%!p___________________________________________________________________
which is expanded with respect to the functions $F_a^5$, $F_a$ and
$F_a^{\text{tr}}$ with the trace term separated from the antisymmetric
and remaining symmetric part.

This structure of the Compton amplitude is a generic one because it
is also valid for the ordinary non-forward case: The additional
structures arising due to the momentum $k$ are hidden in the summation
over the structure functions $F_a$, $F_a^5$ and $F_a^{\text{tr}}$ and
in the definitions of $\PU$ and $\QU$. The reason for that result is
an outcome of the twist structure of the operator which is the same
for all the matrix elements under consideration. It holds also for the
case of $n$ outgoing mesons.

For $\PU = p_+ z_+ + p_- z_-$ and summation over the Dirac- and
Pauli-structures one reproduces the form of the Compton amplitude
given in Ref.~\cite{BR00}. Here, the additional terms containing the
functions $F_a^{\text{tr}}$ arise, because the trace term in
(\ref{twi_mat}) has been taken into account. If $k$ is not present in
the process this term vanishes in the limit
%!p___________________________________________________________________
\begin{eqnarray}
 p_i p_j
&\approx&
 0\ ,
\\
 \gamma^\mu p_{1\mu} \; u\kln{p_1}
&\approx&
 0\ ,
\\
 \bar u\kln{p_2} \, \gamma^\mu p_{2\mu}
&\approx&
 0\ ,
\end{eqnarray}
%!p___________________________________________________________________
setting $M$ and $t = (p_2-p_1)^2$ to zero. Therefore these terms
have been omitted in Ref.~\cite{BR00}. If the five kinematic factors
$\KI_a(x,{\mathbf p})$ containing the momentum $k$ are present this is
a priori no longer the case since non-vanishing contractions like
$(\bar u \Fe{k} u)$ appear. Therefore, this term has been taken into
full account here.

In general, terms containing the product $\PU_\rho \KI^{a\rho}$ do
not vanish in the massless limit, while terms containing $\PU^2$,
${\mathbf{p}}_i {\mathbf{p}}_j$, are small compared to the large
invariants. This leads to the approximation
%!p___________________________________________________________________
\begin{eqnarray}
 \frac{1}{\QU^2 + \im \epsilon}
&\approx&
 \frac{1}{q^2 - q.\PU + \im \epsilon}
\\
\nonumber
&=&
 - \frac{1}{q P_+} \cdot \frac{1}{\xi + (z_+ + \eta z_- + \chi z_k) -
\im \epsilon} \ .
\end{eqnarray}
%!p___________________________________________________________________
We apply these approximations to the Compton amplitude and get a
representation for its twist-2 part in the massless limit and in the
extended Bjorken region, which depends explicitly on the three
variables $z_\pm$ and $z_k$:
%!p___________________________________________________________________
\begin{eqnarray}
\label{Compton}
 T_{\mu\nu}^\twz = - 2 \; \int_{\R^3} \d z_+ \, \d z_- \, \d z_k
&&
 \KLEEo{ - \frac{1}{q P_+} \cdot \frac{1}{\xi + (z_+ + \eta z_- + \chi
z_k) - \im \epsilon}}
\\
\nonumber
&&
 \qquad \times \; \KLSo{\KLe{ g_{\mu\rho}\kln{q_\nu - \PU_\nu} +
g_{\nu\rho}\kln{q_\mu - \PU_\mu} - g_{\mu\nu}\kln{q_\rho - \PU_\rho}}
\cdot F_a\kln{z_+,z_-,z_k} } \; \KI^{a\rho}
\\
\nonumber
&&
 \qquad \qquad \oKLS{ + \; g_{\mu\nu} \PU_\rho \cdot
F_a^{\text{tr}}\kln{z_+,z_-,z_k} \; \KI^{a\rho}
 - \, \im \, {\varepsilon_{\mu\nu}}^{\alpha\beta}
\;q_\alpha \, g_{\beta\rho} \cdot
F_a^5\kln{z_+,z_-,z_k}\;\KI^{5\,a\rho} }
\\
\nonumber
&&
 + \; \frac{1}{\kln{q P_+}^2} \cdot \frac{1}{\kln{\xi + (z_+ + \eta
z_- + \chi z_k) - \im \epsilon}^2} \quad
\\
\nonumber
&&
 \qquad \times \; \KLSo{\KLe{ q_\mu \PU_\nu + q_\nu \PU_\mu - \PU_\mu
\PU_\nu - g_{\mu\nu} \; q.\PU }\; \QU_\rho \cdot F_a\kln{z_+,z_-,z_k}
\; \KI^{a\rho} } \;
\\
\nonumber
&&
 \qquad \qquad + \kln{ 2 \, \QU_\mu \QU_\nu - g_{\mu\nu} \, q^2 +
g_{\mu\nu} \; q.\PU } \PU_\rho \; \cdot
F_a^{\text{tr}}\kln{z_+,z_-,z_k}\; \KI^{a\rho}
\\
\nonumber
&&
 \oKLEE{ \qquad \qquad \oKLS{ - \, \im \,
{\varepsilon_{\mu\nu}}^{\alpha\beta} \; q_\alpha \PU_\beta \, \QU_\rho
\; \cdot F_a^5\kln{z_+,z_-,z_k} \; \KI^{5\,a\rho} } }~.
\end{eqnarray}
%!p___________________________________________________________________
This form of the Compton amplitude will be used in the further
considerations.

%%%%%%%%%%%%%%%%%%%%%%%%%%%%%%%%%%%%%%%%%%%%%%%%%%%%%%%%%%%%%%%%%%%%%%
\section{Integral relations}
\label{int_relations}
\setcounter{equation}{0}
%%%%%%%%%%%%%%%%%%%%%%%%%%%%%%%%%%%%%%%%%%%%%%%%%%%%%%%%%%%%%%%%%%%%%%

\noindent
The variables $z_\pm$ and $z_k$ are not directly measurable because
they appear as Fourier variables of the distribution amplitudes $f_a$.
The scaling variable $\xi$, however, can be regarded as a physical
quantity and it is therefore quite natural to use the new variable
%!p___________________________________________________________________
\begin{equation}
 t \equiv z_+ + \eta z_- + \chi z_k
\end{equation}
%!p___________________________________________________________________
as integration variable in the denominators of Eq.~(\ref{Compton}).

By the definition (\ref{PU}) the vector $\PU$ contains the variables
$z_\pm$ and $z_k$ and must be rewritten using the set
$\kls{t,z_-,z_k}$. This is simply done by introducing the combinations
%!p___________________________________________________________________
\begin{eqnarray}
 \pi
&\equiv&
 P_- - \eta P_+\ ,
\\
 \tilde \pi
&\equiv&
 k - \chi P_+ \ ,
\end{eqnarray}
%!p___________________________________________________________________
which leads to the representation
%!p___________________________________________________________________
\begin{equation}
\label{neues_PU}
 \PU = P_+ \, t + \pi \, z_- + \tilde \pi \, z_k \ .
\end{equation}
%!p___________________________________________________________________
The 4--vectors $\pi$ and $\tilde{\pi}$ define off--collinear
directions w.r.t. to the direction $P_+$; vectors along this momentum
are denoted as collinear. Note that these vectors are non-forward
still, since $\eta \neq 0$. $\PU$ contains collinear and off-collinear
contributions, the former of which are associated with the scaling
variable $t$ only. It will turn out that these collinear parts play
the dominant role in the process considered.

Different powers of $\PU$ contained in (\ref{Compton}), of course,
will lead to a whole series of structure functions corresponding to
different moments in $z_-$ and $z_k$. Therefore, let us generally
define double moments of the triple-valued distribution amplitudes
leading to {\em single-valued distributions} as follows:
%!p___________________________________________________________________
\begin{eqnarray}
\label{neue_sture}
 \hat F_{n_1 n_2}^a\kln{t;\eta,\chi}
&\equiv&
 \frac{1}{t^{n_1+n_2}} \int \d z_- \! \int \d z_k \;\; z_-^{n_1} \,
z_k^{n_2} \; F^a\kln{t - \eta z_- - \chi z_k, \, z_-, \, z_k}
\\
\nonumber
&=&
 \frac{1}{t^{n_1+n_2}} \int_0^1 \frac{\d\tau}{\tau} \;\; \tau^{n_1 +
n_2} \cdot \hat f_{n_1 n_2}^a\kln{\frac{t}{\tau}; \, \eta,\, \chi}
\\
&=&
 \int_t^{\text{sign}\kln{t}} \frac{\d\lambda}{\lambda} \;\;
\lambda^{-n_1 - n_2} \cdot \hat f_{n_1 n_2}^a \kln{\lambda; \, \eta,\,
\chi}
\end{eqnarray}
%!p___________________________________________________________________
with
%!p___________________________________________________________________
\begin{equation}
 \hat f_{n_1 n_2}^a\kln{\frac{t}{\tau}; \, \eta,\, \chi} \, \equiv
\int \d z_- \! \int \d z_k \;\; {z}_{-}^{n_1} \, {z}_{k}^{n_2} \, \hat
f^a\kln{\frac{t}{\tau} - \eta z_- - \chi z_k, \, z_-, \, z_k} .
\end{equation}
%!p___________________________________________________________________
In fact, the values $n_i = 0,1$ occur in the antisymmetric part and
the values $n_i = 0,1,2$ in the symmetric part. To keep the discussion
short, we consider the antisymmetric part of the Compton amplitude in
full detail and give the result for the symmetric part only for the
leading terms, i.e.,~suppressing the trace terms. The explicit
calculation shows that the latter terms do not contribute to the
leading order in $\nu$.

Because the contractions $\PU_\rho \KI^{a\rho}$ are only present for
the Dirac structure, $\KI^{1\rho} = (\bar u\,\gamma^\rho \,u)$, we
discuss this term separately. Then, the antisymmetric part of
(\ref{Compton}) is given by
%!p___________________________________________________________________
\begin{eqnarray}
 T_{\kle{\mu\nu}}^\twz
&\approx&
 2\im \; {\varepsilon_{\mu\nu}}^{\alpha\beta} \; \frac{q_\alpha}{q
P_+} \int_{\R^3} \d z_+ \, \d z_- \, \d z_k
\kle{\frac{g_{\beta\rho}}{\xi+t-\im \epsilon} - \frac{q_\rho}{q P_+}
\frac{\PU_\beta}{\kln{\xi+t - \im \epsilon}^2}} F_a^5(z_+,z_-,z_k) \,
\KI^{5\,a\rho}({\mathbf p})
\\
\nonumber
&=&
 2\im \; {\varepsilon_{\mu\nu}}^{\alpha\beta} \; \frac{q_\alpha}{q P_+
} \int_{-1}^1 \d t \kle{\frac{g_{\beta\rho} \; \hat F_{00}^{5a}(t) }
{\xi+t - \im \epsilon} - \frac{q_\rho}{q P_+} \,
\frac{t\kln{P_{\!+\beta} \hat F_{00}^{5a}(t) + \pi_\beta \hat
F_{10}^{5a}(t) + \tilde \pi_\beta \hat F_{01}^{5a}(t) }} { \kln{\xi+t
- \im \epsilon}^2}} \KI^{5\,a\rho}({\mathbf p}) \, .
\end{eqnarray}
%!p___________________________________________________________________
At this point it is necessary to perform a partial integration in
the second term proportional to $q_\rho$ using the formula
%!p___________________________________________________________________
\begin{equation}
\label{part_int}
 \int_{-1}^1 \d t \frac{t^m}{\kln{\xi+t - \im \epsilon}^2} \cdot \hat
F_{n_1 n_2}^{5a}\kln{t;\eta,\chi} = \int_{-1}^1 \d t
\frac{t^{m-1}}{\xi+t - \im \epsilon} \KLE{m \, \cdot \hat F_{n_1
n_2}^{5a}\kln{t;\eta,\chi} - t^{-n_1-n_2} \cdot \hat f_{n_1
n_2}^{5a}\kln{t;\eta,\chi} } \;\; ,
\end{equation}
%!p___________________________________________________________________
which leads to
%!p___________________________________________________________________
\begin{eqnarray}
\label{comppol}
 T_{\kle{\mu\nu}}^\twz
&\approx&
 2\im \, {\varepsilon_{\mu\nu}}^{\alpha\beta} \; \frac{q_\alpha}{q
P_+}\int_{-1}^1 \d t\;\frac{1}{\xi+t-\im\epsilon}
\\
\nonumber
&&
 \qquad \times \; \kle{g_{\beta\rho} \, \hat F_{00}^{5a} -
\frac{q_\rho}{q P_+ } \kls{P_{+\beta}\kln{\hat F_{00}^{5a}-\hat
f_{00}^{5a}}+\pi_\beta\kln{\hat F_{10}^{5a}-\frac{\hat f_{10}^{5a}}{t}
} + \tilde\pi_\beta\kln{\hat F_{01}^{5a}-\frac{\hat f_{01}^{5a}}{t} }
} } \KI^{5\,a\rho} \; .
\end{eqnarray}
%!p___________________________________________________________________
Several tensor structures contribute. Note that in the foregoing
discussion {\it no} assumption has been made on the direction of the
nucleon spin. As the polarization of the initial state nucleons in
experiment is performed in outer magnetic fields, the direction of the
nucleon spin is not related to other vectors in the system except for
the condition $S_i.p_i = 0$ to hold.

The form (\ref{comppol}) of the polarized Compton amplitude is very
interesting because it includes a Wandzura-Wilczek (WW) like relation
between the distribution amplitudes being associated to two of the
tensor structures~\footnote{For special choices of the nucleon spin
vector, e.g. when directed along the momentum $P_+$, the second tensor structure may
become identical to the former one up to sub--leading corrections of
${\mathcal{O}}(M^2/\nu)$ as well--known from the case of forward scattering, see
e.g. Ref.~\cite{BLTA}. Although the second structure appears as
kinematically suppressed it is still present and a WW like relation between
the twist--2 contributions of the two amplitudes exists. For other
choices of the spin vector this suppression does not occur.}. This
relation becomes obvious, once the definitions
%!p___________________________________________________________________
\begin{eqnarray}
 \G_1^a\kln{t;\eta,\chi}
&\equiv&
 \hat f_{00}^{5a}\kln{t;\eta,\chi}
\\
\label{5.11}
 \G_2^a\kln{t;\eta,\chi}
&\equiv&
 - \hat f_{00}^{5a}\kln{t;\eta,\chi} + \int_t^{\text{sign}\kln{t}}
\frac{\d\lambda}{\lambda} \; \hat f_{00}^{5a}\kln{\lambda;\eta,\chi}
\\
\label{5.14}
 \G_3^a\kln{t;\eta,\chi}
&\equiv&
 - \frac{\hat f_{10}^{5a}\kln{t;\eta,\chi} }{t} +
\int_t^{\text{sign}\kln{t}} \frac{\d\lambda}{\lambda^2} \; \hat
f_{10}^{5a}\kln{\lambda;\eta,\chi}
\\
\label{5.13}
 \G_4^a\kln{t;\eta,\chi}
&\equiv&
 - \frac{\hat f_{01}^{5a}\kln{t;\eta,\chi} }{t} +
\int_t^{\text{sign}\kln{t}} \frac{\d\lambda}{\lambda^2} \; \hat
f_{01}^{5a}\kln{\lambda;\eta,\chi}
\end{eqnarray}
%!p___________________________________________________________________
are made. This leads to a very simple form of the antisymmetric part
of the Compton amplitude, namely,
%!p___________________________________________________________________
\begin{equation*}
 T_{\kle{\mu\nu}}^\twz = 2\im \, {\epsilon_{\mu\nu}}^{\alpha\beta}\;
\frac{q_\alpha}{q P_+ } \int_{-1}^1 \d t \; \frac{1}{\xi+t - \im
\epsilon} \KLEo{g_{\beta\rho}\Kln{\G_1^a + \G_2^a} } \oKLE{ -
\frac{q_\rho}{q P_+ } \Kln{P_{\!+\beta}\,\G_2^a + \pi_\beta\,\G_3^a +
\tilde \pi_\beta\,\G_4^a}} \KI^{5\,a\rho} \ .
\end{equation*}
%!p___________________________________________________________________
All the above functions ${\sf G}^a_k$ are expectation values of
twist--2 operators.
By definition $\G_1^a$ and $\G_2^a$ obey the following integral
relation
%!p___________________________________________________________________
\begin{equation}
\label{WW_relation}
 \G_2^a\kln{t;\eta,\chi} = - \G_1^a\kln{t;\eta,\chi} +
\int_t^{\text{sign}\kln{t}} \frac{\d\lambda}{\lambda} \;
\G_1^a\kln{\lambda;\eta,\chi} \;\; .
\end{equation}
%!p___________________________________________________________________
This relation between two twist--2 quantities
can be viewed as a generalization of the WW-relation known
from forward scattering \cite{WW77},\footnote{Needless to say that
the Lorentz decomposition of the complete Compton amplitude delivers
higher twist corrections to the quantities ${\sf G}^a_k$ as well;
in some cases twist--3 terms are present in the massless limit.
These contributions are not dealt with in the present paper. We mention
that in the case when quark and target mass corrections are taken into
account all contributions may receive higher twist corrections starting
with twist--3, as has been shown for the forward
case in Ref.~\cite{BLTA}. Integral relation between higher twist
contributions do also exist.}.
It is a property of the collinear
part of the polarized Compton amplitude $T_{\kle{\mu\nu}}^\twz$ and
connected to the $00$-moment in $z_-$ and $z_k$. The off-collinear
vectors $\pi$ and $\tilde \pi$ are connected to the new parton
distributions $\G_3^a$ and $\G_4^a$, which obey an integral
representation in terms of the functions $\hat{f}^{5a}_{10}$ and
$\hat{f}^{5a}_{01}$ respectively, similar to $\G_2^a$. However, unlike
the case for $\G_1$ the respective functions do not appear in the
Compton amplitude. Therefore we obtain at the present level only one
Wandzura--Wilczek like relation between the twist--two parts of the
respective amplitudes. Of course all the functions $\G_i$ do receive
higher twist contributions, which were not discussed in the present
paper. Also these contributions as emerging in the different
amplitudes may obey similar integral relations. The generalization
(\ref{WW_relation}) of the WW-relation has been obtained in
Ref.~\cite{BR00} for the ordinary non-forward process
(\ref{ohne_meson}) without outgoing meson. The foregoing discussion
shows that it remains valid for the more general process
(\ref{mit_meson}).

The Wandzura--Wilczek relation Eq.~(\ref{WW_relation}) is an
identity between physical amplitudes which emerges at the level of
{\em geometric} twist--2 as a consequence of the fact that the
distribution $\hat{f}^{5a}_{00}(t,\eta;\chi)$ appears two times in the
decomposition of the Compton amplitude. These relations which
determine the (geometric) twist-2 content of the dynamical twist-3
distributions have to be called {\em geometric WW relations}
\cite{BLKO,BLTA,BR00,BL01,L01a}. They are obtained by using solely
group theoretical means lying behind the definition of (genuine)
geometric twist, Eq.~(\ref{twist}). These WW-relations have to be
distinguished from the so-called {\it dynamical} WW-relations being
obtained by using the QCD equations of motion as has been done by
\cite{dWW}. To bring it to the point:
Geometric WW relations are written
for distributions having equal geometric twist, whereas dynamic WW
relations are written between distributions of equal dynamic twist.
Despite having the same formal structure their physical content is
different. For a detailed discussion of these relations
in the case of meson wave functions see Ref.~\cite{BL01}
which, however, with appropriate modifications also
holds for general non-forward amplitudes 
(See also Ref.~\cite{GL00a} where the case of quark distributions
is discussed).

We now return to the discussion of the term containing the
contractions $\PU_\rho \KI^{a\rho}$. To begin with, we first remind
that only the Dirac structure $\KI^{1\rho}$ is of relevance in this
case and one may use the approximation
%!p___________________________________________________________________
\begin{equation}
 \PU_\rho \KI^{(5)\,1\rho} \approx k_\rho\KLe{t+ \kln{1-\eta}z_- +
\kln{1-\chi} z_k} \, \KI^{(5)\,1\rho}
\end{equation}
%!p___________________________________________________________________
in the limit of vanishing nucleon masses. Even more general, one can
state that the product $q_\rho \KI^{(5)\,1\rho}$ is of order $\nu$ and
$\PU_\rho \KI^{(5)\,a\rho}$ is of order $\mu$ with $\mu \ll \nu$. The
additional terms are therefore of non-leading type, but are
interesting because these structures arise due to the meson momentum
$k$.

Inserting the above approximation and performing again partial
integrations leads to the complete form of $T_{\kle{\mu\nu}}^\twz$.
%!p___________________________________________________________________
\begin{eqnarray}
\label{extrastructur}
 T_{\kle{\mu\nu}}^\twz
&=&
 2\im \, {\varepsilon_{\mu\nu}}^{\alpha\beta}\; \frac{q_\alpha}{q P_+
} \int_{-1}^1 \d t \; \frac{1}{\xi+t - \im \epsilon} \KLSSo{ } \KLE{
g_{\beta\rho}\Kln{\G_1^a + \G_2^a} \! - \frac{q_\rho}{q P_+} \Kln{
P_{\!+\beta}\,\G_2^a + \pi_\beta \, \G_3^a + \tilde\pi_\beta \,\G_4^a}
} \KI^{5\,a\rho}
\\
\nonumber
&&
 \quad \qquad \qquad \quad \qquad \qquad \quad \qquad \qquad \oKLSS {
+ \kle{\frac{k_\rho}{q P_+ } \Kln{P_{\!+\beta} \, \G'_2 + \pi_\beta \,
\G'_3 + \tilde \pi_\beta \, \G'_4}} \KI^{5\,1\rho} }
\end{eqnarray}
%!p___________________________________________________________________
with
%!p___________________________________________________________________
\begin{eqnarray}
\label{5.18}
 \G'_2
&=&
 t \kle{\kln{\hat F_{00}^{51} - \frac{\hat f_{00}^{51} }{2}} +
\kln{1-\eta} \kln{\hat F_{10}^{51} - \frac{\hat f_{10}^{51}}{2t} } +
\kln{1-\chi} \kln{\hat F_{01}^{51} - \frac{\hat f_{01}^{51}}{2t} } }
\\
\label{5.19}
 \G'_3
&=&
 t \kle{\kln{\hat F_{10}^{51} - \frac{\hat f_{10}^{51} }{2t}} +
\kln{1-\eta} \kln{\hat F_{20}^{51} - \frac{\hat f_{20}^{51}}{2t^2} } +
\kln{1-\chi} \kln{\hat F_{11}^{51} - \frac{\hat f_{11}^{51}}{2t^2} } }
\\
\label{5.20}
 \G'_4
&=&
 t \kle{\kln{\hat F_{01}^{51} - \frac{\hat f_{01}^{51} }{2t}} +
\kln{1-\eta} \kln{\hat F_{11}^{51} - \frac{\hat f_{11}^{51}}{2t^2} } +
\kln{1-\chi} \kln{\hat F_{02}^{51} - \frac{\hat f_{02}^{51}}{2t^2} } }
\; \; .
\end{eqnarray}
%!p___________________________________________________________________
The meson momentum $k$ and the momentum transfer $q$ are connected
to similar structures, but the functions $\G'_i$ related to $k$ are
far more complicated than $\G_i$. However, despite that fact the
expressions within the parentheses in Eqs.~(\ref{5.18}) --
(\ref{5.20}) could be written in the same manner as the expressions
(\ref{5.11}) -- (\ref{5.13}) showing {\it potential} WW-like
expressions for the twist-2 distributions
$\hat f^{5a}_{n_1n_2} (t;\eta,\chi)$.

For the symmetric part $T_{\{\mu\nu\}}^\twz$ of the Compton
amplitude we perform the same calculational steps as in the
antisymmetric part, namely:
%!p<><><><><><><><><><><><><><><><><><><><><><><><><><><><><><><><><><
\begin{itemize}
%!p<><><><><><><><><><><><><><><><><><><><><><><><><><><><><><><><><><
\item
{Separation of $\PU_\rho \KI^{a\rho}-$contractions; this also
includes the trace terms.}
%!p<><><><><><><><><><><><><><><><><><><><><><><><><><><><><><><><><><
\item
{Insertion of $\PU = P_+ \, t + \pi \, z_- + \tilde \pi \, z_k $
into the Compton amplitude (\ref{Compton}).}
%!p<><><><><><><><><><><><><><><><><><><><><><><><><><><><><><><><><><
\item
{Definition of the structure functions $\hat F_{n_1 n_2}^a$ using
(\ref{neue_sture}).}
%!p<><><><><><><><><><><><><><><><><><><><><><><><><><><><><><><><><><
\item
{Partial integration with respect to $t$ by formula
(\ref{part_int}).}
%!p<><><><><><><><><><><><><><><><><><><><><><><><><><><><><><><><><><
\item
{Projection onto the collinear part.}
%!p<><><><><><><><><><><><><><><><><><><><><><><><><><><><><><><><><><
\end{itemize}
%!p<><><><><><><><><><><><><><><><><><><><><><><><><><><><><><><><><><
After partial integration the symmetric part of the Compton
amplitude has the following form:
%!p___________________________________________________________________
\begin{eqnarray}
 T_{\kls{\mu\nu}}^\twz \approx - 2 \int_{-1}^1 \d t \;\; \frac{1}{\xi
+ t - \im \epsilon} \; \;
&&
 \frac{1}{q P_+} \KLSSo{ g_{\mu\nu} \; q_\rho \; \hat f_{00}^a -
\Kln{g_{\mu\rho} q_\nu + g_{\nu\rho} q_\mu } \, \hat F_{00}^a +
\Kln{g_{\mu\rho} P_{\!+\nu} + g_{\nu\rho} P_{\!+\mu} } \,t\, \hat
F_{00}^a }
\\
\nonumber
&&
 \qquad \quad + \Kln{g_{\mu\rho} \pi_\nu + g_{\nu\rho} \pi_\mu } \,t\,
\hat F_{10}^a + \Kln{g_{\mu\rho} \tilde\pi_\nu + g_{\nu\rho}
\tilde\pi_\mu } \,t\, \hat F_{01}^a
\\
\nonumber
 +
&&
 \frac{q_\rho}{q P_+} \KLEo{\Kln{q_\mu P_{\!+\nu} + q_\nu P_{\!+\mu}
}\kln{ \hat F_{00}^a - \hat f_{00}^a } - P_{\!+\mu} P_{\!+\nu} \kln{
2t \, \hat F_{00}^a - t\,\hat f_{00}^a } }
\\
\nonumber
&&
 \qquad \quad + \Kln{q_\mu \pi_\nu + q_\nu \pi_\mu} \kln{\hat F_{10}^a
- \hat f_{10}^a } - \Kln{P_{\!+\mu}\pi_\nu + P_{\!+\nu}\pi_\mu}
\kln{t\, \hat F_{10}^a - t\,\hat f_{10}^a }
\\
\nonumber
&&
 \qquad \quad + \Kln{q_\mu \tilde\pi_\nu + q_\nu \tilde\pi_\mu}
\kln{\hat F_{01}^a - \hat f_{01}^a } - \Kln{P_{\!+\mu}\tilde\pi_\nu +
P_{\!+\nu}\tilde\pi_\mu} \kln{t\,\hat F_{01}^a - t\,\hat f_{01}^a }
\\
\nonumber
&&
 \qquad \quad - \pi_\mu \pi_\nu \kln{ 2t \, \hat F_{20}^a - t \, \hat
f_{20}^a } - \tilde\pi_\mu \tilde\pi_\nu \kln{2t \, \hat F_{02}^a - t
\, \hat f_{02}^a}
\\
\nonumber
&&
 \qquad \quad \oKLSS{ \oKLE { - \Kln{\pi_\mu \tilde \pi_\nu + \pi_\nu
\tilde\pi_\mu} \kln{2t \, \hat F_{11}^a - t \, \hat f_{11}^a } } } \,
\KI^{a \rho} \; .
\end{eqnarray}
%!p___________________________________________________________________
The collinear part of $T_{\kls{\mu\nu}}^\twz$, which contains only
functions $\hat F_{00}^a$ and $\hat f_{00}^a$, can be written as
%!p___________________________________________________________________
\begin{eqnarray}
\label{callan_gross}
 T_{\kls{\mu\nu}}^\twz = - 2 \int_{-1}^1 \d t \;\; \frac{1}{\xi + t -
\im \epsilon} \; \frac{1}{q P_+} \!\!
&\KLSSo{&
 q_\rho \kln{g_{\mu\nu} - \frac{q_\mu P_{\!+\nu} + q_\nu P_{\!+\mu}
}{q P_+}} \hat f_{00}^a\kln{t;\eta,\chi} + \frac{q_\rho}{2 \, q P_+}
\; P_{\!+\mu} P_{\!+\nu} \; 2t \, \hat f_{00}^a\kln{t;\eta,\chi} }
\\
\nonumber
&&
 - \kle{ q_\mu \kln{g_{\nu\rho} - P_{\!+\nu} \frac{q_\rho}{q P_+} } +
q_\nu \kln{g_{\mu\rho} - P_{\!+\mu} \frac{q_\rho}{q P_+} } } \hat
F_{00}^a\kln{t;\eta,\chi}
\\
\nonumber
&&
 \oKLSS{ + \kle{ P_{\!+\mu} \kln{g_{\nu\rho} - P_{\!+\nu}
\frac{q_\rho}{q P_+} } + P_{\!+\nu} \kln{g_{\mu\rho} - P_{\!+\mu}
\frac{q_\rho}{q P_+} } } \; t \, \hat F_{00}^a\kln{t;\eta,\chi} } \;
\KI^{a\rho} \; .
\end{eqnarray}
%!p___________________________________________________________________
Looking at the tensor structure of (\ref{callan_gross}), the
functions
%!p___________________________________________________________________
\begin{eqnarray}
 \F^a_1\kln{t;\eta,\chi}
&=&
 \hat f_{00}^a \kln{t;\eta,\chi}\ ,
\\
 \F^a_2\kln{t;\eta,\chi}
&=&
 2t \cdot \hat f_{00}^a \kln{t;\eta,\chi}\ ,
\end{eqnarray}
%!p___________________________________________________________________
appear as natural structure functions and obey a Callan-Gross like
relation \cite{CG69}:
%!p___________________________________________________________________
\begin{equation}
 \F^a_2\kln{t;\eta,\chi} = 2t \cdot \F^a_1\kln{t;\eta,\chi}\ .
\end{equation}
%!p___________________________________________________________________
Similar to Ref.~\cite{BR00} one observes that the remainder
contributions in Eq.~(\ref{callan_gross}) are suppressed for large
values of $\nu$, a property of the off-collinear terms. One may see
this, contracting the structures ${\cal K}^{a\rho}$ with the
respective tensors in front. As well known from forward scattering,
the Callan--Gross relation receives corrections both from higher
orders in the coupling constant and due to mass effects, see
e.g. Ref.~\cite{GP}, and therefore as well in the non-forward case. On the
other hand, the Wandzura--Wilczek relation for geometric twist--2
turns out to be rigidly stable, see e.g. Ref.~\cite{BLTA}.

Including also $\PU_\rho \KI^{a\rho}$-contractions in the
calculation will result in an additional tensor structure analogous to
(\ref{extrastructur}) with $q_\rho$ replaced by $k_\rho$ in
Eq.~(\ref{callan_gross}). The corresponding structure functions will
not be given in explicit form.

%%%%%%%%%%%%%%%%%%%%%%%%%%%%%%%%%%%%%%%%%%%%%%%%%%%%%%%%%%%%%%%%%%%%%%
\section{Evolution equations for the distribution amplitudes}
\label{evo}
\setcounter{equation}{0}
%%%%%%%%%%%%%%%%%%%%%%%%%%%%%%%%%%%%%%%%%%%%%%%%%%%%%%%%%%%%%%%%%%%%%%

\noindent
The scaling violations of the operator matrix elements and
distribution amplitudes of the process considered are described by the
renormalization group equations governing the ultra--violet behavior
of the light--cone operators. The corresponding equations for the
distribution amplitudes $f_a\kln{z_+,z_-,z_k}$ are called evolution
equations, which we are going to discuss in $x-$ and $z-$space. Since
the flavor content of the operators (\ref{str_opra}) and
(\ref{str_oprb}) has been suppressed in the preceding sections, we
treat only the flavor non-singlet evolution equations as an example.
The singlet evolution equations are of quite similar structure (see
e.g. Ref.~\cite{BGR99}; for earlier work see Refs.~\cite{BGR87, BB89}).

The non-singlet renormalization group equation for the twist-2
vector operator reads
%!p___________________________________________________________________
\begin{eqnarray}
 \mu^2 \frac{\d}{\d\mu^2} \,
O_\beta^{\twz}\kln{\kappa_1\tx,\kappa_2\tx;\mu^2} &=& \int_{\R^2} \!
\D^2\!\kappa' \;\, \gamma \kln{\kappa_1,\kappa_2; \kappa_1',
\kappa_2'} O_\beta^{\twz} \kln{\kappa_1'\tx, \kappa_2' \tx; \mu^2} \,
.
\end{eqnarray}
%!p___________________________________________________________________
By contraction with $\tx^\beta$ it is obvious that the scalar
twist-2 operator $O^{\twz} = \tx^\beta O_\beta^\twz$ obeys exactly the
same renormalization group equation because multiplication with
$\tx^\beta$ commutes with the differentiation on the left and with the
integration on the right hand side. This gives the renormalization
group equation for the scalar operator which on the light cone already
is of twist-2:
%!p___________________________________________________________________
\begin{eqnarray}
\label{ren_sca}
 \mu^2 \frac{\d}{\d\mu^2} \, O\kln{\kappa_1\tx,\kappa_2\tx;\mu^2} &=&
\int_{\R^2} \! \D^2\!\kappa' \;\, \gamma \kln{\kappa_1,\kappa_2;
\kappa_1', \kappa_2'} O\kln{\kappa_1'\tx, \kappa_2' \tx; \mu^2} \, .
\end{eqnarray}
%!p___________________________________________________________________
In the last equations the integration measure
%!p___________________________________________________________________
\begin{eqnarray}
\label{ren_mes}
 \D^2\!\kappa' &\equiv& \d\kappa_1' \, \d\kappa_2' \;
\theta\kln{\kappa_1 - \kappa_1'} \theta\kln{\kappa_1' - \kappa_2} \,
\theta\kln{\kappa_1 - \kappa_2'} \theta\kln{\kappa_2' - \kappa_2}
\end{eqnarray}
%!p___________________________________________________________________
has been introduced. In Refs.~\cite{BGR87,BGR99} it is shown that
the non-local anomalous dimension matrix $\gamma$ is invariant under
translations and scale transformations,
%!p___________________________________________________________________
\begin{eqnarray}
 \gamma\kln{\kappa_1,\kappa_2 \, ; \, \kappa_1',\kappa_2' }
&=&
 \gamma\kln{\kappa_1 - \kappa_0,\kappa_2 - \kappa_0 \, ; \, \kappa_1'
- \kappa_0,\kappa_2' - \kappa_0 }
\\
\nonumber
&=&
 \lambda^2 \, \gamma\kln{\lambda\kappa_1,\lambda\kappa_2 \, ; \,
\lambda\kappa_1',\lambda\kappa_2' } \ ,
\end{eqnarray}
%!p___________________________________________________________________
which reduces the number of independent variables of $\gamma$ by
two. By first changing the variables from $\kappa_{1,2}$ to
$\kappa_\pm$, followed by a translation by $\kappa_+$ and a scaling by
$\kappa^{-1}$ one derives the following form of the evolution kernel
$\gamma$:
%!p___________________________________________________________________
\begin{eqnarray}
\label{ren_mas}
 \gamma\kln{\kappa_1,\kappa_2 \, ; \, \kappa_1',\kappa_2' }
&=&
 \tilde\gamma\kln{\kappa_+,\kappa \, ; \, \kappa_+',\kappa' }
\\
\nonumber
&=&
 \tilde\gamma\kln{0,\kappa \, ; \, \kappa_+' - \kappa_+,\kappa' }
\\
\nonumber
&=&
 \frac{1}{\kappa^2 } \; \tilde\gamma\kln{0,1 \, ; \, \frac{\kappa_+' -
\kappa_+}{\kappa},\frac{\kappa'}{\kappa} }
\\
\nonumber
&\equiv&
 \frac{1}{\kappa^2 } \; \widetilde K\kln{w_1,w_2}
\end{eqnarray}
%!p___________________________________________________________________
with
%!p___________________________________________________________________
\begin{equation}
 w_1 = \frac{\kappa_+' - \kappa_+}{\kappa} \qquad \text{and} \qquad
w_2 = \frac{\kappa'}{\kappa} \, .
\end{equation}
%!p___________________________________________________________________
The variables $\kappa_i'$ and $w_i$ are connected by the following
transformation
%!p___________________________________________________________________
\begin{equation}
 \kln{\begin{array}{c}{\kappa_1'} \\ {\kappa_2'} \end{array}} =
\kln{\begin{array}{cc} {\kappa} & - {\kappa} \\ {\kappa} & {\kappa}
\end{array}} \kln{\begin{array}{c} w_1 \\ w_2 \end{array}} +
\kln{\begin{array}{c} \kappa_+ \\ \kappa_+ \end{array}}~.
\end{equation}
%!p___________________________________________________________________
It is therefore more natural to use $w_1$ and $w_2$ as integration
variables in the renormalization group equation (\ref{ren_sca})
instead of $\kappa_1'$ and $\kappa_2'$. The integration measures are
related by
%!p___________________________________________________________________
\begin{equation}
\label{ren_tra}
 \D^2 \! \kappa' = \kappa^2 \, \D^2 \! w \; ,
\end{equation}
%!p___________________________________________________________________
where $\D^2\! \kappa'$ and $\D^2 \! w$ include the suitable
$\theta$-functions realizing the integration ranges of $\kappa_i'$
(\ref{ren_mes}) and $w_i$:
%!p___________________________________________________________________
\begin{eqnarray*}
\nonumber
 \D^2 \! w &\equiv& \frac{1}{2} \; \d w_1 \, \d w_2 \; \theta\kln{1 +
w_1 - w_2} \, \theta\kln{1 - w_1 + w_2} \, \theta\kln{1 + w_1 + w_2}
\, \theta\kln{1 - w_1 - w_2}
\end{eqnarray*}
%!p___________________________________________________________________
The measure $\D^2 \! w$ can be divided into two parts, because
$\widetilde K\kln{w_1,w_2}$ under the exchange
$w_1 \lra - w_1,\,w_2 \lra - w_2$ obeys the following relations, cf.
Ref.~\cite{BGR99}~:
%!p___________________________________________________________________
\begin{equation}
 \widetilde K\kln{w_1,w_2}= \widetilde K\kln{-w_1,w_2}= -\widetilde
K\kln{w_1,-w_2}\ .
\end{equation}
%!p___________________________________________________________________
Putting (\ref{ren_mas}) and (\ref{ren_tra}) into the renormalization
group equation for the scalar operator results in
%!p___________________________________________________________________
\begin{eqnarray}
\label{renorm_group}
 \mu^2 \frac{\d}{\d\mu^2} \, O\kln{\kappa_1\tx,\kappa_2\tx;\mu^2} &=&
\int_{\R^2} \! \D^2\!w \;\, \widetilde K\kln{w_1,w_2} \;
O\kln{\kappa_1'\tx, \kappa_2' \tx; \mu^2} \; .
\end{eqnarray}
%!p___________________________________________________________________
For the explicit structure of $\widetilde K\kln{w_1,w_2}$
see Refs.~\cite{BGR99, BGR87, BB89}.
The equation (\ref{renorm_group}) will now be considered for the
matrix elements of the
scalar operator which are, according to equation (\ref{mat_zerl}),
given by
%!p___________________________________________________________________
\begin{eqnarray}
 \Matel{p_2,k}{O\kln{ \kappa_1 \tx, \kappa_2 \tx, \mu^2}}{p_1} &=&
\e^{\im \kappa_+ \, \tx P_-} \, \KI^a\kln{\tx,p_2,k,p_1}\, \tilde
f_a\kln{\kappa \tx P_+ , \kappa \tx P_-, \kappa \tx k, \mu^2} \, .
\end{eqnarray}
%!p___________________________________________________________________
>From this one obtains directly the evolution equation for the
distribution amplitudes $\tilde f_a$ in $x$-space:
%!p___________________________________________________________________
\begin{eqnarray}
\label{ren_xra}
 \mu^2 \frac{\d}{\d\mu^2} \; \tilde f_a\kln{\kappa\tx P_+ , \kappa\tx
P_-, \kappa\tx k; \mu^2} &=& \int_{\R^2} \D^2 \! w \;\; \widetilde
K\kln{w_1,w_2}\; \e^{\im w_1\kappa \, \tx P_- } \; \, \tilde
f_a\kln{w_2\kappa \, \tx P_+,w_2 \kappa \, \tx P_-,w_2\kappa \, \tx
k;\mu^2} \, .
\end{eqnarray}
%!p___________________________________________________________________
Because we are interested in evolution equations in $z$-space, we
perform a Fourier transformation of equation (\ref{ren_xra}). The
physically relevant transforms of $\tilde f_a$ are given by
%!p___________________________________________________________________
\begin{eqnarray}
 f_a\kln{z_+,z_-,z_k; \mu^2} &=& \int_\R \frac{\d \! \kln{\kappa \tx
P_+}}{2\pi} \int_\R \frac{\d \! \kln{\kappa \tx P_-}}{2\pi} \int_\R
\frac{\d \! \kln{\kappa \tx k}}{2\pi} \;\e^{\im \kappa\tx \kln{P_+ z_+
+ P_-z_- + kz_k}} \; \tilde f_a\kln{\kappa \tx P_+, \kappa \tx P_-,
\kappa \tx k; \mu^2} \, .
\end{eqnarray}
%!p___________________________________________________________________
Carrying out these transformations one arrives at the following
result:
%!p___________________________________________________________________
\begin{eqnarray}
\nonumber
 \mu^2 \frac{\d}{\d\mu^2} \; f_a\kln{z_+,z_-,z_k, \mu^2}
&=&
 \int_\R \frac{\d \! \kln{\kappa \tx P_+}}{2\pi} \int_\R \frac{\d \!
\kln{\kappa \tx P_-}}{2\pi} \int_\R \frac{\d \! \kln{\kappa \tx
k}}{2\pi} \; \e^{\im \kappa\tx \kln{P_+ z_+ + P_-z_- + kz_k}} \;
\int_{\R^2} \! \D^2 \! w \; \widetilde K\kln{w_1,w_2}
\\
\nonumber
&&
 \qquad \times \; \int_{\R^3} \! \D^3 \! z' \;\;
f_a\kln{z'_+,z'_-,z'_k;\mu^2} \;\; \e^{-\im w_2\kappa\tx \kln{P_+ z'_+
+ P_- z'_- + kz'_k}} \; \e^{\im w_1 \kappa \tx P_- }
\\
\nonumber
&&
\\
\nonumber
&=&
 \int_{\R^2} \! \D^2 \! w \; \widetilde K\kln{w_1,w_2}
\\
\nonumber
&&
 \qquad \times \;\int_{\R^3} \! \D^3 \! z' \;
f_a\kln{z'_+,z'_-,z'_k;\mu^2} \; \delta\kln{z_+ - w_2 z'_+} \;
\delta\kln{z_- - w_2 z'_- + w_1} \; \delta\kln{z_k - w_2 z'_k}
\\
\nonumber
&&
\\
&=&
 \int_{\R^2} \! \D^2 \! z' \;\; \frac{1}{\abs{z_+}} \; \widetilde
K\kln{z'_-\frac{z_+}{z'_+}-z_- \, , \, \frac{z_+}{z'_+} } \;
f_a\kln{z'_+,z'_-,z_k\frac{z'_+}{z_+};\mu^2}~.
\end{eqnarray}
%!p___________________________________________________________________
with $\D^3 \! z'$ given by (\ref{ddreiz}) and $\D^2 \! z'$ defined by
%!p___________________________________________________________________
\begin{eqnarray}
\nonumber
 \D^2 \! z' &=& 2 \, \d z_+' \, \d z_-' \; \theta(1-z_+' + z_-') \,
\theta(1+z_+' - z_-') \, \theta(1-z_+' - z_-') \, \theta(1+z_+' + z_-')
\end{eqnarray}
%!p___________________________________________________________________
We introduce the evolution kernel
%!p___________________________________________________________________
\begin{equation}
 \Gamma\kln{z_+, z_-, z_+', z_-'} \equiv \frac{1}{\abs{z_+}} \,
\widetilde K\kln{z'_-\frac{z_+}{z'_+}-z_- \, , \, \frac{z_+}{z'_+} },
\end{equation}
%!p___________________________________________________________________
which leads to the this evolution equation
%!p___________________________________________________________________
\begin{eqnarray}
\label{evolution_equ}
 \mu^2 \frac{\d}{\d\mu^2} \; f_a\kln{z_+,z_-,z_k; \mu^2} &=&
\int_{\R^2} \! \D^2 \! z' \;\, \Gamma\kln{z_+, z_-, z_+', z_-'}\; f_a \!
\kln{z'_+,z'_-,z_k\,\frac{z'_+}{z_+};\mu^2} .
\end{eqnarray}
%!p___________________________________________________________________
Let us point to the remarkable fact that the variable $z_k$
connected to the meson momentum $k$ only appears as a parameter in
$f_a$ and is not contained in the evolution kernel $\Gamma$. The same
observation has been made in Ref.~\cite{BR01} recently in the case of
diffractive scattering, where the parameters $\eta$ or $x_\PP$ behave
in the same way. In so far some of the scaling variables of a problem,
in the present case the variables $\chi_i$, play another role than
others, as here $\xi$ and $\eta$, which interfere with the evolution.

This evolution equation is a fundamental equation because it
describes the evolution of the triple-valued distribution amplitudes
$f_a\kln{z_+,z_-,z_k}$ in $z-$space. These amplitudes are the basic
objects for the construction of the structure functions $F_a, F_a^5$
and $F_a^{\text{tr}}$ in (\ref{Fa}) and (\ref{Ga}). They are also used
in the definition of the single-valued functions
$\hat F_{n_1 n_2}^a\kln{t;\eta,\chi}$ in (\ref{neue_sture}). The
scaling violations of these functions are obtained solving
Eq.~(\ref{evolution_equ}) and inserting the functions $f_a$ into
Eqs.~(\ref{Fa},\ref{Ga}).

It is also possible to obtain another evolution equation for
$\hat f^a_{00}\kln{t;\eta,\chi}$ in the variable $t$, which is
compatible with the former equation. This single-variable evolution
equation governs the evolution of the structure functions contained in
the collinear part of the Compton amplitude.

To begin with, we first show that the distribution amplitude
$\hat f^a_{00}\kln{t;\eta,\chi}$ given by
%!p___________________________________________________________________
\begin{equation}
\label{representation_f00}
 \hat f^a_{00}\kln{t;\eta,\chi} = \int \d z_- \! \int \d z_k \; \hat
f^a\kln{t - \eta z_- - \chi z_k, \, z_-, \, z_k}
\end{equation}
%!p___________________________________________________________________
has another representation obtained as
%!p___________________________________________________________________
\begin{eqnarray}
\label{representation_b}
 \KI^a({\mathbf p})\,\hat f^a_{00}\kln{t;\eta,\chi} = \int \frac{\d \!
\kln{\kappa \, \tx P_+ }}{2\pi} \; \e^{\im \kappa \,t\, \tx P_+ } \;
\Matel{p_2,k}{O\kln{-\kappa \tx,\kappa \tx}}{p_1} \PIPe{\tx P_- = \eta
\, \tx P_+ \; ; \;\; \tx k = \chi \, \tx P_+ } .
\end{eqnarray}
%!p___________________________________________________________________
The constraints
%!p___________________________________________________________________
\begin{eqnarray}
\label{constraint_a}
 \tx P_-
&=&
 \eta \, \tx P_+
\\
\label{constraint_b}
 \tx k
&=&
 \chi \, \tx P_+
\end{eqnarray}
%!p___________________________________________________________________
appearing in the former equation are the scaling relations
(\ref{scalen_variable}) in $x-$space. Using the representation
(\ref{scalar_matrix_element}) under these constraints leads to the
result (\ref{representation_f00}).

To derive the single-variable evolution equation we form matrix
elements of equation (\ref{renorm_group}),
%!p___________________________________________________________________
\begin{eqnarray}
&&
 \mu^2 \frac{\d}{\d\mu^2} \Matel{p_2,k}{O\kln{-\kappa\tx,\kappa
\tx}}{p_1} \PIPe{\tx P_- = \eta \, \tx P_+ \; ; \;\; \tx k = \chi \,
\tx P_+ }
\\
\nonumber
&&
 \qquad \qquad = \int_{\R^2} \D^2\!w \;\, \widetilde K\kln{w_1,w_2} \;
\e^{\im w_1\kappa\tx P_-} \;
\Matel{p_2,k}{O\kln{-w_2\kappa\tx,w_2\kappa\tx}}{p_1} \PIPe{\tx P_- =
\eta \, \tx P_+ ; \; \tx k = \chi \, \tx P_+ } ,
\end{eqnarray}
%!p___________________________________________________________________
and perform the Fourier transformation in the variable
$\kappa \tx P_+$ according to (\ref{representation_b}). The direct
calculation leads to
%!p___________________________________________________________________
\begin{eqnarray}
 \mu^2 \frac{\d}{\d\mu^2} \hat f^a_{00}\kln{t;\eta,\chi,\mu^2} &=&
\int_{-1}^1 \d t' \; \gamma\kln{t,t';\eta} \; \hat
f^a_{00}\kln{t';\eta,\chi,\mu^2} \, ,
\end{eqnarray}
%!p___________________________________________________________________
with the evolution kernel
%!p___________________________________________________________________
\begin{equation}
 \gamma\kln{t,t';\eta} = \int \d w_2 \; \frac{1}{\abs{\eta}} \,
\widetilde K\kln{\frac{w_2 t' - t}{\eta},w_2} \; .
\end{equation}
%!p___________________________________________________________________
Like the evolution kernel $\Gamma\kln{z_+,z_- ; z_+',z_-'}$, also
$\gamma\kln{t,t';\eta}$ does not depend on any $k-$dependent variables
like $\chi$ or $z_k$ being related to the meson momentum.

%%%%%%%%%%%%%%%%%%%%%%%%%%%%%%%%%%%%%%%%%%%%%%%%%%%%%%%%%%%%%%%%%%%%%%
\section{Generalization to an arbitrary number of outgoing mesons}
\label{gen}
\setcounter{equation}{0}
%%%%%%%%%%%%%%%%%%%%%%%%%%%%%%%%%%%%%%%%%%%%%%%%%%%%%%%%%%%%%%%%%%%%%%

\begin{figure}[ht]\centering
\epsfig{file=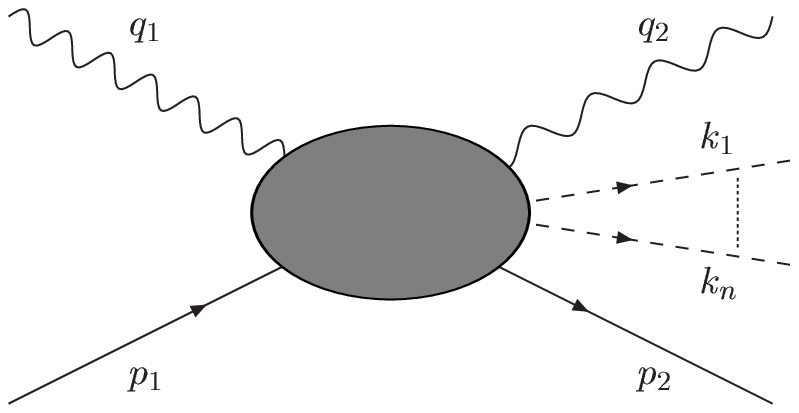}
\caption{Process with $n$ outgoing mesons}
\label{many_mesons}
\end{figure}

\noindent
In this section we summarize the generic properties of the results
obtained in the preceding sections and extend it to an arbitrary
number of outgoing mesons. Generally, one may state that all the above
results remain valid under slight modifications if two or more
outgoing scalar mesons are present in the process,
%!p___________________________________________________________________
\begin{equation}
\label{mit_vielen_mesonen}
 \gamma_1^*(q_1) + {\rm H}(p_1) \Ra \gamma_2^*(q_2) + {\rm H}(p_2) +
{\rm M}(k_1) + \dots + {\rm M}(k_n)\ ,
\end{equation}
%!p___________________________________________________________________
as shown in Fig.~\ref{many_mesons}.
To fix the kinematic domain of this process, all meson momenta $k_i$
are connected to different scaling variables $\chi_i$ defined by
%!p___________________________________________________________________
\begin{equation}
 \chi_i = \frac{k_i}{q P_+} \ ,
\end{equation}
%!p___________________________________________________________________
where $P_+$ and $P_-$ are obviously given by
%!p___________________________________________________________________
\begin{eqnarray}
 P_\pm = \Big(p_2 + \sum_{i=1}^{n} k_i\Big) \pm p_1 \ ,
\end{eqnarray}
%!p___________________________________________________________________
and $\xi$ and $\eta$ are introduced as in (\ref{scalen_variable}).
In order to compute the twist-2 part of the Compton amplitude
%!p___________________________________________________________________
\begin{equation}
 T_{\mu\nu}\kln{p_2,k_1,\dots,k_n,p_1,q} = \im \ix \; \e^{\im qx}
\matel{p_2,S_2;k_1,\dots,k_n} {R\,T\kle{J_\mu\kln{\frac{x}{2}}J_\nu
\kln{-\frac{x}{2}}{\mathcal{S}}} }{p_1,S_1}
\end{equation}
%!p___________________________________________________________________
for the general process (\ref{mit_vielen_mesonen}) one applies the
same approximations to the operator $\hat T_{\mu\nu}$ as in
Sections.~\ref{str} and \ref{twi}. Thus one uses the approximation
(\ref{str_ope}) and applies the twist-2 projection (\ref{twi_tw2}).

The technique used in Section~\ref{twi} to construct the matrix
elements $\matel{p_2,k}{O^{(5)}}{p_1}$ can be carried out for an
arbitrary number $n$ of scalar mesons where the additional meson
momenta $\kls{k_2, ..., k_n}$ enlarge the set of kinematic factors
$\KI^a(\tx, {\mathbf p})$. However, these factors are easy to guess
and their number is given by
%!p___________________________________________________________________
\begin{equation}
 N_{\text{scalar}} = \klso{\begin{array}{lll} 2^{n+1} + \binom{n+3}{4}
& {\rm for} &n\geq 1 \\ 2 & {\rm for} & n=0 \end{array} } \ ,
\end{equation}
%!p___________________________________________________________________
for the scalar matrix element. This formula also reproduces the
number of kinematic factors in the ordinary non-forward case: the
Dirac- and Pauli-structures.

Having all kinematic factors $\KI^a$ at hand, one introduces the
related structure functions $\tilde f_a$ and writes down the following
decomposition of the scalar matrix element
%!p___________________________________________________________________
\begin{eqnarray}
\nonumber
 \Matel{p_2,k_1,\dots,k_k}{O\kln{-\kappa \tx, \kappa \tx}}{p_1} &=&
\sum_{a=1}^{N_{\text{scalar}}} \KI_a\kln{\tx,p_2,k_1,\dots,k_n,p_1}\;
\tilde f_a\kln{\kappa \tx P_+,\,\kappa \tx P_-,\, \kappa \tx k_1,\dots
,\kappa \tx k_n,\, {\mathbf{p}}_i{\mathbf{p}}_j;\,\mu^2 } \, .
\end{eqnarray}
%!p___________________________________________________________________
With this representation one goes through the same steps of the
calculation as in the preceding sections, namely
%!p<><><><><><><><><><><><><><><><><><><><><><><><><><><><><><><><><><
\begin{itemize}
%!p<><><><><><><><><><><><><><><><><><><><><><><><><><><><><><><><><><
\item
{Fourier transformation of $\tilde f_a$}
%!p<><><><><><><><><><><><><><><><><><><><><><><><><><><><><><><><><><
\item
{Application of the twist-2 projector}
%!p<><><><><><><><><><><><><><><><><><><><><><><><><><><><><><><><><><
\item
{Computation of the Compton amplitude.}
%!p<><><><><><><><><><><><><><><><><><><><><><><><><><><><><><><><><><
\end{itemize}
%!p<><><><><><><><><><><><><><><><><><><><><><><><><><><><><><><><><><
The result is again of the form (\ref{Compton_b}) with a larger
$z-$space and $\PU$ given by
%!p___________________________________________________________________
\begin{equation}
 \PU = P_+ \, z_+ + P_- \, z_- + \sum_{i=1}^n \chi_i \, z_{k_i} \ .
\end{equation}
%!p___________________________________________________________________
In this sense, the form (\ref{Compton_b}) of the Compton amplitude
is a generic result holding for a large class of processes. The more
explicit form (\ref{Compton}) is generalized by replacing
$\chi \, z_k$ by $\sum \chi_i \, z_{k_i}$.

It is even possible to interpret the Wandzura--Wilczek and
Callan--Gross relations obtained in Section~\ref{int_relations} as
generic properties of the collinear parts of these processes. Making
the substitution
%!p___________________________________________________________________
\begin{equation}
 \PU = P_+ \, t + \pi \, z_- + \sum_{i=1}^n \tilde \pi_i \, z_{k_i}
\quad \text{with} \quad \tilde \pi_i = k_i - \chi_i P_+ \ ,
\end{equation}
%!p___________________________________________________________________
and projecting onto the collinear part one finds again the relations
%!p___________________________________________________________________
\begin{eqnarray}
 \G_2^a\kln{t;\eta,\chi_1,\dots, \chi_n}
&=&
 - \G_1^a\kln{t;\eta,\chi_1,\dots, \chi_n} +
\int_t^{\text{sign}\kln{t}} \frac{\d  \lambda}{\lambda} \;
\G_1^a\kln{\lambda;\eta,\chi_1,\dots, \chi_n} \, ,
\\
 \F^a_2\kln{t;\eta,\chi_1,\dots, \chi_n}
&=&
 2t \cdot \F^a_1\kln{t;\eta,\chi_1,\dots, \chi_n} .
\end{eqnarray}
%!p___________________________________________________________________
Looking at the derivation of the evolution equation in
Section~\ref{evo}, it is not difficult to find the appropriate
generalization to an arbitrary number of mesons
(${\mathbf k} = (k_1,\dots,k_n)$):
%!p___________________________________________________________________
\begin{eqnarray}
 \mu^2 \frac{\d}{\d\mu^2} \; f_a\kln{z_+,z_-,z_{\mathbf k} ; \mu^2} =
\int_{\R^2} \! \D^2 \! z' \; \frac{\abs{z_+'}^{n-1}}{\abs{z_+}^n} \;
\widetilde K\kln{z'_-\frac{z_+}{z'_+}-z_- \, , \, \frac{z_+}{z'_+} }
\; f_a\kln{z'_+,z'_-,z_{\mathbf k}\frac{z'_+ }{z_+} ; \mu^2}\ .
\end{eqnarray}
%!p___________________________________________________________________
For $n=0$, this general evolution equation is reproducing the flavor
non-singlet part of the evolution equation given in Ref.~\cite{BGR99}
for the ordinary non-forward scattering. As in the case of one meson
in Section~\ref{evo} the variables $z_{k_i}$ connected to the meson
momenta only appear as parameters and do not contribute to the
evolution kernel,
%!p___________________________________________________________________
\begin{equation}
 \Gamma^n\kln{z_+, z_-, z_+', z_-'} \equiv
\frac{\abs{z_+'}^{n-1}}{\abs{z_+}^n} \, \widetilde
K\kln{z'_-\frac{z_+}{z'_+}-z_- \, , \, \frac{z_+}{z'_+} }\ .
\end{equation}
%!p___________________________________________________________________
Only the number of mesons is relevant for the structure of this
kernel. The single-variable evolution equation is of the same form as
in the preceding section. One only has to enlarge the number of
scaling parameters $\chi_i$.
%!p___________________________________________________________________
\begin{eqnarray}
 \mu^2 \frac{\d}{\d\mu^2} \hat
f^a_{00}\kln{t;\eta,\chi_1,\dots,\chi_n,\mu^2} &=& \int_{-1}^1 \d t'
\; \gamma\kln{t,t';\eta} \; \hat
f^a_{00}\kln{t';\eta,\chi_1,\dots,\chi_n,\mu^2} \, .
\end{eqnarray}
%!p___________________________________________________________________

%%%%%%%%%%%%%%%%%%%%%%%%%%%%%%%%%%%%%%%%%%%%%%%%%%%%%%%%%%%%%%%%%%%%%%
\section{Conclusions}
%%%%%%%%%%%%%%%%%%%%%%%%%%%%%%%%%%%%%%%%%%%%%%%%%%%%%%%%%%%%%%%%%%%%%%

\noindent
We studied the structure of the virtual Compton amplitude for
deep-inelastic non-forward scattering
${\gamma^*}(q_1) + {\mathrm H}(p_1)
\Ra {\gamma^*}(q_2) + {\mathrm H}(p_2) + {\mathrm M}(k)$
at the level of the twist--2 contributions in lowest order in QCD in
the massless limit. In the extended Bjorken region, i.e.,
$\{(qP_+), -q^2\} \ra \infty$ with $-q^2/(qP_+), (qP_-)/(qP_+)$ and
$(kP_-)/(qP_+)$ kept fixed, the twist-2 contributions to the Compton
amplitude were calculated using the non-local operator product
expansion for general spin states. In this approximation the Compton
amplitude consists of five kinematically independent parts which in
the limit $k \ra 0$ reduce to the well known Dirac- and Pauli-type
amplitudes. A decomposition of the Compton amplitude was performed
with respect to the helicity states of both (virtual) photons. In
complete analogy the (electromagnetic) gauge invariance of the
non-local light-cone expansion holds at the level of the $S$--matrix
since the fact that the leptonic currents are conserved. Due to this,
only those contributions in the Compton amplitude are projected out,
which obey gauge invariance. Integral relations generalizing the
Callan-Gross and Wandzura-Wilczek relations for unpolarized and
polarized forward-scattering are derived by reduction to the collinear
parts of the Compton amplitude and, thereby, reducing the
triple-valued distribution amplitudes to one-valued ones (for
$(qP_-)/(qP_+)$ and $(kP_-)/(qP_+)$ fixed). In this connection
attention has been drawn to the difference between geometric and
dynamic WW-relations being related to different notions of twist. The
evolution kernels of these distribution amplitudes are obtained from
the (well-known) non-local anomalous dimensions of the (scalar)
twist-2 light-ray operators $O^{\twz}(\kappa_1\tx, \kappa_2\tx)$ and
$O^{5 \, \twz}(\kappa_1\tx, \kappa_2\tx)$; they are independent of the
meson momentum $k$. These results show that deeply virtual Compton
scattering off nucleons in the case of additional meson production
behaves quite similar to the case where mesons are absent. Both the
basic structural relations as well as the scaling violations are the
same in both cases. However, the structure of the Compton amplitude is
different in general, however only with corrections of
${\mathcal{O}}(1/\sqrt{\nu})$ or less.

\acknowledgments
\noindent
The authors are grateful to M. Lazar and M. Diehl for various useful
discussions. In addition, J. Eilers gratefully acknowledges the Graduate
College "Quantum field theory" at Center for Theoretical Studies of
Leipzig University for financial support.

%!p###################################################################
\begin{appendix}
%%%%%%%%%%%%%%%%%%%%%%%%%%%%%%%%%%%%%%%%%%%%%%%%%%%%%%%%%%%%%%%%%%%%%%
\section{Projection of the twist-2 operator onto the light-cone}
\label{Projection}
\setcounter{equation}{0}
\setcounter{section}{8}
%%%%%%%%%%%%%%%%%%%%%%%%%%%%%%%%%%%%%%%%%%%%%%%%%%%%%%%%%%%%%%%%%%%%%%

\noindent
This appendix is devoted to the derivation of the results
(\ref{twi_enda}) and (\ref{twi_endb}) from the off--cone twist-2
non-local quark-antiquark operators obtained in Ref.~\cite{GLR01}.

As has been shown there the nonlocal quark-antiquark operator of
geometric twist-2 has the following structure
$\kln{\kln{\bar\psi\gamma_\mu\psi}(q) =
\int \d^4 \! x \;  \exp\{ \im\,(qx)\} \bar\psi(x)\gamma_\mu\psi(-x)}$:
%!p___________________________________________________________________
\begin{align}
\label{nl_O2}
 O^{\rm tw\,2}_\alpha (x,-x)
&
 =\int\!\frac{\d^4 \! q}{(2\pi)^4}\,
\big(\bar{\psi}\gamma_\mu\psi\big)(q) \left(2+q\pd_q\right)\int_0^1\d
\tau \bigg\{ \Big[ \left(3+q\pd_q\right)\delta^\mu_{\alpha} -\im \tau
q^\mu x_\alpha\Big] {\cal H}_2(q|\tau x)
\\
\nonumber
&
 \qquad\qquad +\Big[ \left(3+q\pd_q\right) \left(
\left(4+q\pd_q\right)\im t q_\alpha x^\mu
-\hbox{\large$\frac{1}{2}$}(\im \tau)^2 \big(q^2 x^\mu x_\alpha + x^2
q^\mu q_\alpha \big) \right) +\hbox{\large$\frac{1}{4}$} (\im t)^3
q^\mu q^2 x_{\alpha}x^2 \Big] {\cal H}_3(q|\tau x) \bigg\}.
\end{align}
%!p___________________________________________________________________
Its $n-$th moment is given by
%!p___________________________________________________________________
\begin{align}
\label{XO2}
\nonumber
 O^{\text{tw\,2}}_{\alpha n}(x)
&
 = \frac{1}{(n+1)^2} \int\!\frac{\d^4 \! q}{(2\pi)^4}\,
\big(\bar{\psi}\gamma_\mu\psi\big)(q) \bigg\{ \delta_\alpha^\mu
\,h_n^2(q|x) - q^\mu x_\alpha \,h_{n-1}^2(q|x)
\\
&
 \qquad\qquad\qquad +2 x^\mu q_\alpha \,h_{n-1}^3(q|x) -\big(x^\mu
x_\alpha q^2+q^\mu q_\alpha x^2\big) \,h_{n-2}^3(q|x)
+\hbox{\large$\frac{1}{2}$}\, q^\mu x_\alpha x^2 q^2 \,h_{n-3}^3(q|x)
\bigg\}.
\end{align}
%!p___________________________________________________________________
Here, for notational simplicity we used the following abbreviations:
%!p___________________________________________________________________
\begin{align}
\label{H_nonloc}
 {\cal H}_\nu(q|x)
&
 = \sqrt{\pi} \left(\sqrt{(qx)^2-q^2 x^2}\right)^{1/2-\nu} \;
J_{\nu-1/2}\left(\hbox{\large$\frac{1}{2}$} \sqrt{(qx)^2-q^2
x^2}\right) \e^{\im qx/2},
\\
\label{h_loc}
 h_n^\nu(q|x)
&
 = \left(\hbox{\large$\frac{1}{2}$}\sqrt{q^2 x^2}\right)^n
\!C_n^\nu\bigg(\frac{qx}{\sqrt{q^2 x^2}}\bigg).
\end{align}
%!p___________________________________________________________________
The relation between the non-local and the local operators,
Eqs.~(\ref{nl_O2}) and (\ref{XO2}), off the light-cone is obtained by
observing that the Bessel functions are generating functions of the
Gegenbauer polynomials (see, e.g.,~Ref.~\cite{PBM}, Eq.~II.5.13.1.3):
%!p___________________________________________________________________
\begin{align}
 \sum_{n=0}^\infty\frac{a^n}{(2\nu)_n}\, C^\nu_n(z) & =
\Gamma\left(\nu+\hbox{\large$\frac{1}{2}$}\right)
\left(\frac{a}{2}\sqrt{1-z^2}\right)^{1/2-\nu} \;
J_{\nu-1/2}\left(a\sqrt{1-z^2}\right) \e^{za},
\end{align}
%!p___________________________________________________________________
where
$(2\nu)_n = 2\nu (2\nu +1) \ldots (2\nu+n-1) = \Gamma(n+2\nu)/\Gamma(2\nu)$
is the Pochhammer symbol.

The projection onto the light-cone is obtained most easily by first
considering the local operators. Because of the series expansion of
the Gegenbauer polynomials (see, e.g.,~Ref.~\cite{PBM}, Appendix
II.11),
%!p___________________________________________________________________
\begin{align}
\label{GB10}
 C_n^\nu(z)=\frac{1}{(\nu-1)!}\sum_{k=0}^{[\frac{n}{2}]}
\frac{(-1)^k(n-k+\nu-1)!}{k!(n-2 k)!}\,(2z)^{n-2k},
\end{align}
%!p___________________________________________________________________
one observes that from the expression (\ref{h_loc}) on the
light-cone, $x^2=0$, only the term with the highest power, i.e., for
$k=0$, survives:
%!p___________________________________________________________________
\begin{equation}
 h_n^\nu(q|\tx) = \frac{(n+\nu-1)!}{n!(\nu-1)!}(q\tx)^n\ .
\end{equation}
%!p___________________________________________________________________
Using these results in the expression (\ref{XO2}) we obtain:
%!p___________________________________________________________________
\begin{align}
\label{xo2}
\nonumber
 O^{\text{tw\,2}}_{\alpha n}(\tx)
&
 =\; \frac{1}{n+1}\; \int\!\frac{\d^4 \! q}{(2\pi)^4}\,
\big(\bar{\psi}\gamma_\mu\psi\big)(q) \Big\{ \delta_\alpha^\mu
\,(q\tx)^n + \tx^\mu q_\alpha \,n\,(q\tx)^{n-1}\Big\}
\\
&
 - \frac{1}{(n+1)^2} \int\!\frac{\d^4 \! q}{(2\pi)^4}\,
\big(\bar{\psi}\gamma_\mu\psi\big)(q) \tx_\alpha\,\Big\{ q^\mu
\,n\,(q\tx)^{n-1} +\frac{1}{2}\tx^\mu \,q^2 \, n\,(n-1)\,(q\tx)^{n-2}
\Big\}.
\end{align}
%!p___________________________________________________________________
Now, using
%!p___________________________________________________________________
\begin{equation}
 \frac{1}{n+1}= \int_0^1 \d\tau \; \tau^n \quad{\text{and}}\quad
\frac{1}{(n+1)^2}= - \int_0^1 \d\tau \; \tau^n \ln\tau
\end{equation}
%!p___________________________________________________________________
we are able to re-sum over $n$ according to
%!p___________________________________________________________________
\begin{align}
 O^{\rm tw\,2}_\alpha (-\kappa\tx,\kappa\tx) = \sum_{n=0}^\infty
\frac{(-\im\kappa)^n}{n!} O^{\text{tw\,2}}_{\alpha n}(\tx)\ .
\end{align}
%!p___________________________________________________________________
There are two options of doing this. In the first instance we may
replace $\im\kappa q_\mu$ in Eq.~(\ref{xo2}) by the derivative
$\tilde\pd_\mu$ acting on the exponential $\exp\{\im\kappa(q\tx)\}$.
This way one retains the expression (\ref{TW2}) of the non-local
twist-2 operator on the light-cone from which the expression
(\ref{twi_enda}) has been derived. On the other hand, after taking
matrix elements of (\ref{nl_O2}) and observing the definitions
(\ref{Fa}) and (\ref{Ga}) of the distribution amplitudes
$F_a({\mathbf z})$ and $F_a^{\rm tr}({\mathbf z})$, one obtains
exactly the expression (\ref{twi_enda}). Analogous results hold for
the axial vector case (\ref{twi_endb}).

The same result could have been obtained also using the Poisson
integral for the Bessel functions (cf.,~Ref.~\cite{BE},
Eq.~II.7.12.7),
%!p___________________________________________________________________
\begin{align}
 \Gamma(\nu + \hbox{\large$\frac{1}{2}$}) J_\nu(z) & =
\frac{1}{\sqrt\pi} \Big(\frac{z}{2}\Big)^\nu \int^1_{-1} \d
t\,(1-t^2)^{\nu-1/2}\, \e^{\im tz}
\end{align}
%!p___________________________________________________________________
for ${\rm Re~} \nu > - \hbox{\large$\frac{1}{2}$}$, in order to
express the functions (\ref{H_nonloc}) on the light-cone by
%!p___________________________________________________________________
\begin{equation}
 {\cal H}_\nu(q|\tx) = \frac{1}{\Gamma(\nu)}
\int^1_{0}\d\lambda\,[\,\lambda\,(1-\lambda)\,]^{\nu-1}\,
\e^{\im\,\lambda\,(q\tx)}\ .
\end{equation}
%!p___________________________________________________________________
Then, after shifting the homogeneous derivations $q\pd_q$ in the
expression (\ref{nl_O2}) to the right and interpreting it as
$\lambda\pd_\lambda$ acting on the exponential, some partial
integrations with respect to $\lambda$ can be performed which,
finally, lead again to the expression (\ref{TW2}) and
(\ref{twi_enda}), respectively.

%%%%%%%%%%%%%%%%%%%%%%%%%%%%%%%%%%%%%%%%%%%%%%%%%%%%%%%%%%%%%%%%%%%%%%
\setcounter{section}{1}
\section{Helicity projections and current conservation}
\label{helicity projection}
\setcounter{section}{9}
\setcounter{equation}{0}
%%%%%%%%%%%%%%%%%%%%%%%%%%%%%%%%%%%%%%%%%%%%%%%%%%%%%%%%%%%%%%%%%%%%%%

In this appendix we construct the helicity projections of the
Compton amplitude generalizing the results of Ref.~\cite{BR00} to the
present case. We start with the construction of the helicity basis of
the two virtual photons $\gamma_1^*$ and $\gamma_2^*$. To simplify
this construction, we choose the Breit frame, in which the relevant
momenta read:
%!p___________________________________________________________________
\begin{eqnarray}
 P_+
&=&
 \kln{\M; \vec 0}\ ,
\\
 P_-
&=&
 \kln{0;0,0, P_{-3}}\ ,
\\
 k
&=&
 \kln{k_0; k_1, k_2, k_3}\ ,
\\
 q
&=&
 \kln{q_0; q_1,0, q_3} \; ,
\end{eqnarray}
%!p___________________________________________________________________
where $\M$ is introduced as a mass scale of the hadronic momentum
$P_+$ with $P_+^2 = \M^2$. To define the helicity basis we introduce
the two reference vectors
%!p___________________________________________________________________
\begin{eqnarray}
 n_0
&=&
 \kln{1; 0,0,0}\ ,
\\
 n_2
&=&
 \kln{0; 0,1,0} \ .
\end{eqnarray}
%!p___________________________________________________________________
The polarization vectors of the photons $\gamma_1^*$ and
$\gamma_2^*$ are then given by
%!p___________________________________________________________________
\begin{eqnarray}
 \varepsilon_{0\mu}^1
&=&
 \frac{1}{N_{01}} \; q_{1\mu}\ ,
\\
 \varepsilon_{0\mu}^2
&=&
 \frac{1}{N_{02}} \; q_{2\mu}\ ,
\\
\nonumber
&&
\\
 \varepsilon_{1\mu}^1 = \varepsilon_{1\mu}^2
&=&
 n_{2\mu}\ ,
\\
 \varepsilon_{2\mu}^1
&=&
 \frac{1}{N_{21}} \; \varepsilon_{\mu\alpha\beta\gamma} \, n_0^\alpha
\, n_2^\beta \, q_1^\gamma\ ,
\\
 \varepsilon_{2\mu}^2
&=&
 \frac{1}{N_{22}} \; \varepsilon_{\mu\alpha\beta\gamma} \, n_0^\alpha
\, n_2^\beta \, q_2^\gamma\ ,
\\
\nonumber
&&
\\
 \varepsilon_{3\mu}^1
&=&
 \frac{1}{N_{31}} \; \kln{ q_{1\mu} \; q_1.n_0 - n_{0\mu} \; q_1.q_1
}\ ,
\\
 \varepsilon_{3\mu}^2
&=&
 \frac{1}{N_{32}} \; \kln{ q_{2\mu} \; q_2.n_0 - n_{0\mu} \; q_2.q_2
}\ .
\end{eqnarray}
%!p___________________________________________________________________
The normalization factors are given by
%!p___________________________________________________________________
\begin{eqnarray}
 N_{01}
&=&
 \nu^{1/2} \sqrt{ \abs{\xi-\eta - \frac{P_-^2}{4\nu}} }\ ,
\\
 N_{02}
&=&
 \nu^{1/2} \sqrt{ \abs{\xi+\eta - \frac{P_-^2}{4\nu}} }\ ,
\\
 N_{21}
&=&
 \frac{\nu}{\mu} \sqrt{\abs{1+ \frac{\mu^2}{\nu} \kln{\xi-\eta} -
\frac{\mu^2 \, P_-^2}{4 \nu^2} }}\ ,
\\
 N_{22}
&=&
 \frac{\nu}{\mu} \sqrt{\abs{1+ \frac{\mu^2}{\nu} \kln{\xi+\eta} -
\frac{\mu^2 \, P_-^2}{4 \nu^2} }}\ ,
\\
 N_{31}
&=&
 \frac{\nu^{3/2}}{\mu} \sqrt{\abs{\xi-\eta + \frac{\mu^2}{\nu}
\kln{\xi- \eta}^2 - P_-^2 \kln{ \frac{1}{4 \nu} - \frac{\mu^2}{2
\nu^2}\kln{\xi - \eta} + \frac{\mu^2 P_-^2}{16 \, \nu^3} } } }\ ,
\\
 N_{32}
&=&
 \frac{\nu^{3/2}}{\mu} \sqrt{\abs{\xi+\eta + \frac{\mu^2}{\nu}
\kln{\xi+\eta}^2 - P_-^2 \kln{ \frac{1}{4 \nu} - \frac{\mu^2}{2
\nu^2}\kln{\xi + \eta} + \frac{\mu^2 P_-^2}{16 \, \nu^3} } } } \ ,
\end{eqnarray}
%!p___________________________________________________________________
which follows from the relations
%!p___________________________________________________________________
\begin{eqnarray}
 (qP_+)
&=&
 \nu\ ,
\\
 (qP_-)
&=&
 \eta \, \nu\ ,
\\
 (qq)
&=&
 - \xi \, \nu\ ,
\\
 (qn_0)
&=&
 \frac{\nu}{\M}\ ,
\end{eqnarray}
%!p___________________________________________________________________
(see definitions (\ref{scalen_variable}) and note that
$n_0 = P_+/\M$ in the Breit frame) and conservation of momentum
%!p___________________________________________________________________
\begin{eqnarray}
 q_1
&=&
 q + \frac{P_-}{2}\ ,
\\
 q_2
&=&
 q - \frac{P_-}{2} \ .
\end{eqnarray}
%!p___________________________________________________________________
The polarization vectors obey the following normalization condition,
%!p___________________________________________________________________
\begin{equation}
 \varepsilon_{a\mu}^i \, \varepsilon_b^{i \mu} = s_a \, \delta_{ab}\ ,
\end{equation}
%!p___________________________________________________________________
with $s_a=-1$ for $a=0,1,2$ and $s_a=1$ for $a=3$. We now use this
helicity basis to compute all matrix elements
%!p___________________________________________________________________
\begin{equation}
 \T_{kl} \equiv \varepsilon_{k}^{2\mu} \, T_{\mu\nu}^\twz \,
\varepsilon_{l}^{1\nu} \qquad \text{with} \qquad k,l \in \kls{0,1,2,3}
\end{equation}
%!p___________________________________________________________________
The result of the straightforward calculation is
%!p___________________________________________________________________
\begin{eqnarray}
\T_{00}^F
&\approx&
 \frac{2}{\nu} \; \frac{1}{\sqrt{\abs{\xi^2 - \eta^2}}} \int_{\R^3} \! \D^3\!z
\; F_a\kln{z_+,z_-,z_k} \; q_\rho \, \KI^{a\rho}
\\
 \T_{11}^F
&\approx&
 2 \int_{\R^3} \!  \D^3\!z \; \kln{ -\frac{1}{\QU^2 + \im \epsilon} +
\frac{q.\PU}{\kln{\QU^2 + \im \epsilon}^2} } F_a \; q_\rho \,
\KI^{a\rho}
\\
 \T_{22}^F
&\approx&
 2 \int_{\R^3} \!  \D^3\!z \; \kln{ -\frac{1}{\QU^2 + \im \epsilon} +
\frac{q.\PU}{\kln{\QU^2 + \im \epsilon}^2} } F_a \; q_\rho \,
\KI^{a\rho}
\\
 \T_{33}^F
&\approx&
 \frac{2}{\nu} \; \frac{1 }{\sqrt{\abs{\xi^2 - \eta^2}} \sqrt{ \abs{ 1
+ \frac{M^2}{\nu}\xi + \frac{M^4}{\nu^2} \kln{\xi^2-\eta^2} } } } \;
\int_{\R^3} \!  \D^3\!z \; F_a\kln{z_+,z_-,z_k} \;q_\rho \, \KI^{a\rho}
\end{eqnarray}
%!p___________________________________________________________________
%!p___________________________________________________________________
\begin{eqnarray}
\nonumber
 \T_{01}^F , \T_{10}^F, \T_{02}^F, \T_{20}^F
&=&
 {\mathcal{O}}\kln{\frac{1}{\sqrt{\nu}}}
\\
 \T_{03}^F
&\approx&
 \frac{2}{\nu} \; \frac{1}{\sqrt{\abs{\xi^2 - \eta^2}} \sqrt{ \abs{ 1+
\frac{M^2}{\nu} \kln{\xi-\eta} } } } \; \int_{\R^3} \!  \D^3\!z \;
F_a\kln{z_+,z_-,z_k} q_\rho \, \KI^{a\rho}
\\
 \T_{30}^F
&\approx&
 \frac{2}{\nu} \; \frac{1}{\sqrt{\abs{\xi^2 - \eta^2}} \sqrt{ \abs{ 1+
\frac{M^2}{\nu} \kln{\xi+\eta} } } } \; \int_{\R^3} \!  \D^3\!z \;
F_a\kln{z_+,z_-,z_k} q_\rho \, \KI^{a\rho}
\\
\nonumber
 \T_{12}^F , \T_{21}^F
&=&
 {\mathcal{O}}\kln{\frac{1}{\nu}}
\\
\nonumber
 \T_{13}^F , \T_{31}^F, \T_{23}^F, \T_{32}^F
&=&
 {\mathcal{O}}\kln{\frac{1}{\sqrt{\nu}}}
\end{eqnarray}
%!p___________________________________________________________________
for the symmetric part. The helicity projections of the trace terms
are all of order $1/\nu$ and therefore not given in explicit form. The
same calculation is is carried out for the antisymmetric part:
%!p___________________________________________________________________
\begin{eqnarray*}
 \T_{00}^{F5} , \T_{11}^{F5}
&=&
 0
\\
 \T_{22}^{F5} , \T_{33}^{F5}
&=&
 {\mathcal{O}}\kln{\frac{1}{\nu}}
\\
 \T_{01}^{F5} , \T_{10}^{F5} , \T_{02}^{F5} , \T_{20}^{F5}
&=&
 {\mathcal{O}}\kln{\frac{1}{\sqrt{\nu}}}
\\
 \T_{03}^{F5} , \T_{30}^{F5}
&=&
 {\mathcal{O}}\kln{\frac{1}{\nu}}
\\
 \T_{13}^{F5} , \T_{31}^{F5} , \T_{23}^{F5} , \T_{32}^{F5}
&=&
 {\mathcal{O}} \kln{\frac{1}{\sqrt{\nu}}}
\end{eqnarray*}
%!p___________________________________________________________________
%!p___________________________________________________________________
\begin{eqnarray}
 \T_{12}^{F5}
&\approx&
 2 \, \int_{\R^3} \!  \D^3\!z \; \kln{ \frac{1}{\QU^2 + \im \epsilon} -
\frac{q.\PU}{\kln{\QU^2 + \im \epsilon}^2} } F^5_a \; q_\rho \,
\KI^{5\,a\rho}
\\
 \T_{21}^{F5}
&\approx&
 2 \, \int_{\R^3} \!  \D^3\!z \; \kln{ - \frac{1}{\QU^2 + \im \epsilon} +
\frac{q.\PU}{\kln{\QU^2 + \im \epsilon}^2} } F^5_a \; q_\rho \,
\KI^{5\,a\rho}
\end{eqnarray}
%!p___________________________________________________________________
Here, we have only kept terms contributing to the highest power in
$\nu$, because all other terms vanish in the limit $\nu\ra\infty$.
Terms proportional to $P_-^2$ in the normalization factors have been
neglected, because they do not contribute to the highest power in
$\nu$. The amplitudes $\T_{00}^F, \T_{33}^F$ and
$\T_{03}^F, \T_{30}^F$ are a priori not of order $1/\nu$, because
$q_\rho \KI^{a\rho}$ is of order $\nu$. But since the integrals
%!p___________________________________________________________________
\begin{equation}
 \int_{\R^3} \!  \D^3\! z \; F\kln{z_+,z_-,z_k} = 0
\end{equation}
%!p___________________________________________________________________
vanish, these amplitudes are also identical to zero.

In the above only the contractions of the helicity vectors with the
Compton amplitude were considered. For the physical process, however,
the corresponding projections for the leptonic tensors
$L_{\mu\nu}^{1,2}$ have to be considered as well to see, which terms
contribute to the physical $S$--matrix. Due to the fact that
%!p___________________________________________________________________
\begin{equation}
 L_{\mu\nu}^1 q_1^\nu = L_{\mu\nu}^2 q_2^\nu \equiv 0~,
\end{equation}
%!p___________________________________________________________________
holds all remaining terms in the projections $\T_{0k}$ and $\T_{k0}$
are annihilated.

In leading order in $\nu$ only the amplitudes $\T_{11}^F, \T_{22}^F$
and $\T_{12}^{F5}, \T_{21}^{F5}$ give a non-vanishing contribution in
the extended Bjorken region, whereas other terms $\T_{kl}^{(5)}$ for
$k,l=1,2,3$ are suppressed at least in ${\mathcal{O}}(1/\sqrt{\nu})$. The explicit
calculation also shows that
%!p___________________________________________________________________
\begin{equation}
 \T_{11}^F = \T_{22}^F \qquad \text{and} \qquad \T_{12}^{F5} = -
\T_{21}^{F5}
\end{equation}
%!p___________________________________________________________________
in leading order. Only two of the sixteen amplitudes are relevant
for $\nu \rightarrow \infty$. Similar results have been obtained in
Ref.~\cite{DGPR}.
\end{appendix}
%!p###################################################################

%%%%%%%%%%%%%%%%%%%%%%%%%%%%%%%%%%%%%%%%%%%%%%%%%%%%%%%%%%%%%%%%%%%%%%

%%%%%%%%%%%%%%%%%%%%%%%%%%%%%%%%%%%%%%%%%%%%%%%%%%%%%%%%%%%%%%%%%%%%%%

\end{document}